\documentclass[sigconf,10pt]{acmart}
\usepackage{flushend}
\usepackage{soul}
\usepackage{latexsym}
\usepackage{amsmath}
\usepackage{gensymb}
\usepackage{balance}
\usepackage{makecell}
\usepackage{pifont}
\usepackage{xspace}
\usepackage{anyfontsize}
\usepackage[inline]{./trackchanges}
\usepackage{subfigure}
\usepackage{enumitem}
\usepackage{algorithm}
\usepackage{algorithmic}
\usepackage{multicol}

\AtBeginDocument{
  \providecommand\BibTeX{{
    \normalfont B\kern-0.5em{\scshape i\kern-0.25em b}\kern-0.8em\TeX}}}

\newcommand{\zerodisplayskips}{%
  \setlength{\abovedisplayskip}{2pt}%
  \setlength{\belowdisplayskip}{2pt}%
  \setlength{\abovedisplayshortskip}{2pt}%
  \setlength{\belowdisplayshortskip}{2pt}}
\appto{\normalsize}{\zerodisplayskips}
\appto{\small}{\zerodisplayskips}
\appto{\footnotesize}{\zerodisplayskips}

\setitemize{topsep=2pt,parsep=0pt,partopsep=0pt,leftmargin=12pt}


\renewcommand\footnotetextcopyrightpermission[1]{}

\acmYear{2025}\copyrightyear{2025}
\acmConference[ACM MobiCom '25]{The 31st Annual International Conference on Mobile Computing and Networking}{November 4--8, 2025}{Hong Kong, China}
\acmBooktitle{The 31st Annual International Conference on Mobile Computing and Networking (ACM MobiCom '25), November 4--8, 2025, Hong Kong, China}
\acmDOI{10.1145/3680207.3723465}
\acmISBN{979-8-4007-1129-9/25/11}


\addeditor{Huixin} 

\newcommand\figcaption{\def\@captype{figure}\caption}
\newcommand\tabcaption{\def\@captype{table}\caption}
\newcommand{\sysname}{PassiveBLE }
\newcommand{\sysnamenospace}{PassiveBLE}
\newcommand{\fig}{Figure }

\begin{document}

\title{PassiveBLE: Towards Fully Commodity-Compatible BLE Backscatter}

\author{\normalsize
Huixin Dong$^{\dagger}$,  Yijie Wu$^{\dagger}$, Feiyu Li$^{\dagger}$, Wei Kuang$^{\dagger}$, Yuan He$^{\#}$, Qian Zhang$^{\triangle}$, Wei Wang$^{\dagger}$$^{*}$} 
\affiliation{%
  \institution{\normalsize
  $^{\dagger}$Huazhong University of Science and Technology,}
  \institution{\normalsize
   $^{\#}$Tsinghua University, $^{\triangle}$Hong Kong University of Science and Technology}
  \country{}
  {\small
  \{huixin, yijiewu, feiyuli, kuangwei, weiwangw\}@hust.edu.cn, heyuan@tsinghua.edu.cn, qianzh@cse.ust.hk
  }
}
 \renewcommand{\shortauthors}{ H. Dong, Y. Wu, F. Li, W. Kuang, Y. He, Q. Zhang, W. Wang}
\settopmatter{printfolios=true}

\begin{abstract}
Bluetooth Low Energy (BLE) backscatter is a promising candidate for battery-free Internet of Things (IoT) applications. 
Unlike existing commodity-level BLE backscatter systems that only enable one-shot communication through BLE advertising packets, we propose PassiveBLE, a backscatter system that can establish authentic and fully compatible BLE connections on data channels. 
The key enabling techniques include (i) a synchronization circuit that can wake up tags and activate backscatter communications with symbol-level accuracy to facilitate BLE data packet generation; (ii) a distributed coding scheme that offloads the major encoding and processing burdens from tags to the excitation source while achieving high throughput; (iii) a BLE connection scheduler to enable fully compatible BLE connection interactions, including connection establishment, maintenance and termination for multiple backscatter tags.
We prototype PassiveBLE tags with off-the-shelf components and also convert the circuits and control logic into ASIC design sketch, whose power consumptions are 491~$\mu$W and 9.9~$\mu$W, 
respectively. Experimental results demonstrate that PassiveBLE achieves a success rate of over 99.9\% in establishing commodity BLE connections. PassiveBLE also achieves commodity-compatible BLE communication with a high goodput of up to 974~kbps in \textit{LE 2M PHY} mode and 532~kbps in \textit{LE 1M PHY} mode, which is about 63.3$\times$ higher than previous commodity-level BLE backscatter system in the same mode.
\renewcommand{\thefootnote}{\fnsymbol{footnote}}
\footnotetext{$^{*}$Corresponding author: Wei Wang.} 
\end{abstract}
\maketitle
 
\section{Introduction}
\label{sec_intro}
Both industry~\cite{BLEAmbientIoT, AmbientIoT_Market_research, AmbientIoT_Williot, AmbientIoT_BLE_vehicle} and academia~\cite{zhang2020reliable,zhang2021commodity, EAScatter,zhang2017freerider,ma2020joint} advocate the development of ultra-low-power Bluetooth Low Energy (BLE) backscatter to enable more power-efficient wireless communication in various applications, including logistics~\cite{zhang2021commodity, AmbientIoT_Williot, AmbientIoT_Market_research,ayyalasomayajula2018bloc}, smart infrastructures~\cite{zhang2021commodity, BLEAmbientIoT, AmbientIoT_Market_research}, and wearable electronics~\cite{InterScatter,BLEAmbientIoT}. For example, the  BLE chips typically consume 82.7\% of the total power in an earbud~\cite{chatterjee2022clearbuds}. Replacing it with a BLE backscatter would provide hundreds of kbps data rates while extending the battery life by approximately tenfold.

Existing BLE backscatter systems~\cite{zhang2017freerider, EAScatter,zhang2020reliable,zhang2021commodity} can only modulate data on advertising packets for one-shot communication.
While promising, advertising-packet-based communication suffers from limited throughput, high delay, and security vulnerabilities. The inefficient modulation scheme in advertising packets limits the throughput of the commodity-compatible BLE backscatter to $8.4$~kbps~\cite{zhang2021commodity}, while some BLE data collection applications, e.g., in-vehicle sensor networks~\cite{AmbientIoT_BLE_vehicle}, demand $125$~kbps - 1~Mbps goodput.
In addition, Tag activation in these systems relies on packet length detection, resulting in a delay of hundreds of milliseconds. However, some applications, such as vehicle sensor networks, require delays on the order of a few milliseconds.

\begin{figure}
    \centering
    \setlength{\abovecaptionskip}{0.cm}
    \setlength{\belowcaptionskip}{0.cm}
    \includegraphics[width=\linewidth]{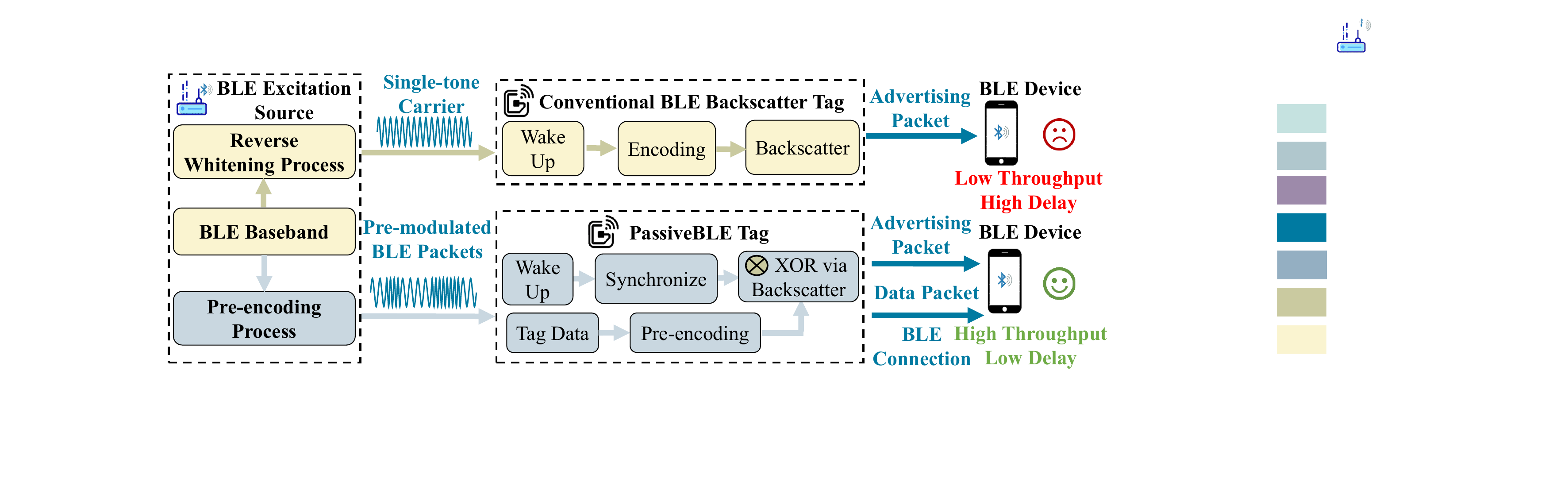}
    \caption{A \sysname tag achieves commodity-compatible BLE connections with both data and advertising packets. In contrast, the traditional commodity-compatible tag~\cite{zhang2021commodity} only employs BLE advertising packets to realize one-shot communication.}
    \label{fig:setup}
\end{figure}

To address the above concerns and push the limits of the BLE backscatter, we seek to answer a key question: \textit{ Can we establish fully compatible and authentic BLE connections between passive backscatter tags and commodity BLE devices with performance comparable to active BLE chips?} 
Upon delving deeper into this problem, we observe that enabling commodity BLE connections requires hardware resources that exceed the capabilities of existing backscatter systems.
This fundamental paradox brings the following three challenges in designing a fully commodity-compatible BLE backscatter.
\begin{itemize}
    \item \textbf{Symbol-level synchronization}. Supporting commodity-compatible BLE connections demands $\mu$s-level latency  (about 150~$\mu$s)~\cite{BLESpecification} and symbol-level accuracy (jitters $\leq 1~\mu$s) to generate data packets. 
    The packet length detection approaches, which are widely used in BLE backscatter systems~\cite{zhang2020reliable,zhang2021commodity,zhang2017freerider,jiang2023dances}, cannot meet the requirement due to the short packet interval in the BLE protocol~\cite{BLESpecification} and insufficient accuracy (e.g., 2~$\mu$s~\cite{zhang2017freerider,zhang2021commodity,zhang2020reliable}) due to envelope variances caused by fading of the wireless channel~\cite{pang2022novel}.
    
    \item \textbf{Tag-affordable BLE channel coding}. Two connected BLE devices need negotiation to acquire dynamic parameters
    e.g., CRC and data whitening parameters, for channel coding~\cite{BLESpecification}, which is unaffordable for backscatter tags. This is because the limited resources on battery-free tags and lack of local oscillators (LO) make it impractical to extract these parameters and encode data with these parameters within several microseconds.
    
    \item \textbf{Compatible BLE connection}. Establishing and maintaining a commodity BLE connection incurs significant storage and processing burdens on parameters negotiation (e.g., channel map exchange) 
    and connection management (e.g., connection establishment, maintenance, and termination). Due to their limited processing capabilities, these power-intensive burdens are unaffordable for tags.
\end{itemize}

To tackle the first challenge, we propose to design a high-accuracy and low-latency synchronization circuit. Unlike conventional backscatter systems that detect the signal envelope~\cite{zhang2020reliable,zhang2021commodity,zhang2017freerider,jiang2023dances} or re-shaped envelope~\cite{huang2024bitalign,guo2022saiyan,na2023leggiero,song2023mumote} for synchronization, we extract the frequency difference between BLE symbols to perform synchronization. Since the frequency of a BLE signal has less variance than amplitude after propagation through a wireless channel~\cite{pang2022novel}, frequency difference extraction-based synchronization can achieve higher accuracy than envelope detection-based methods. Furthermore, since frequency difference extraction based synchronization is realized by detecting the frequency difference of adjacent symbols, rather than detecting the length of the whole packet~\cite{zhang2020reliable,zhang2021commodity,zhang2017freerider}, thereby significantly reducing the synchronization latency. The challenge is that backscatter tags have no LO to extract the symbol frequency difference. 
To address this issue, we leverage the fact that acoustic signals transmit much slower than electromagnetic (EM) signals, and thus use surface acoustic wave (SAW) filters to delay the symbols, and then mix them with non-delayed symbols. In this way, the frequency difference of neighboring symbols is converted to the baseband without LO.      

\begin{figure*}[t]
    \centering
    \setlength{\abovecaptionskip}{0.cm}
    \setlength{\belowcaptionskip}{0.cm}
    \begin{minipage}{0.48\linewidth}
       \includegraphics[width=\linewidth]{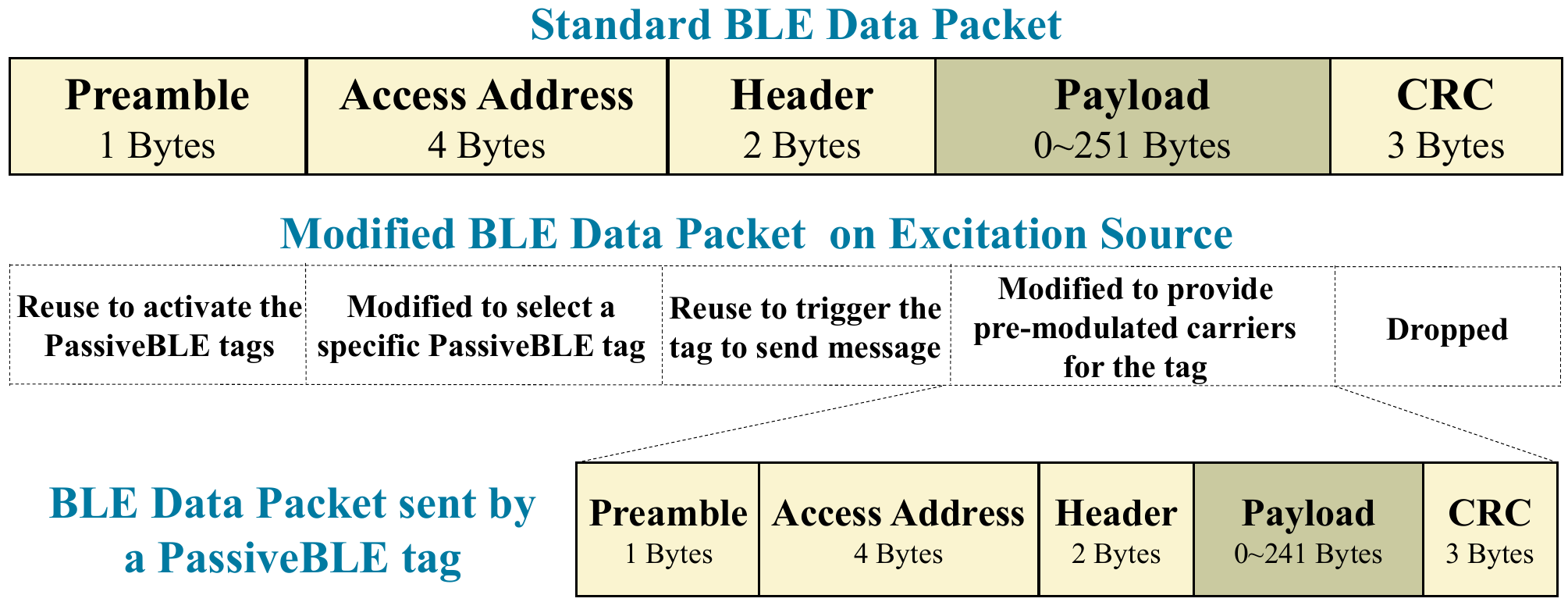}
    \caption{Reuse the structure of BLE data packets to achieve commodity-compatible.}
    \label{fig:Observation}
   \end{minipage}
   \hspace{0.1cm}
    \begin{minipage}{0.50\linewidth}
           \subfigure[Traditional BLE Connection.]{
			\includegraphics[width=0.44\linewidth]{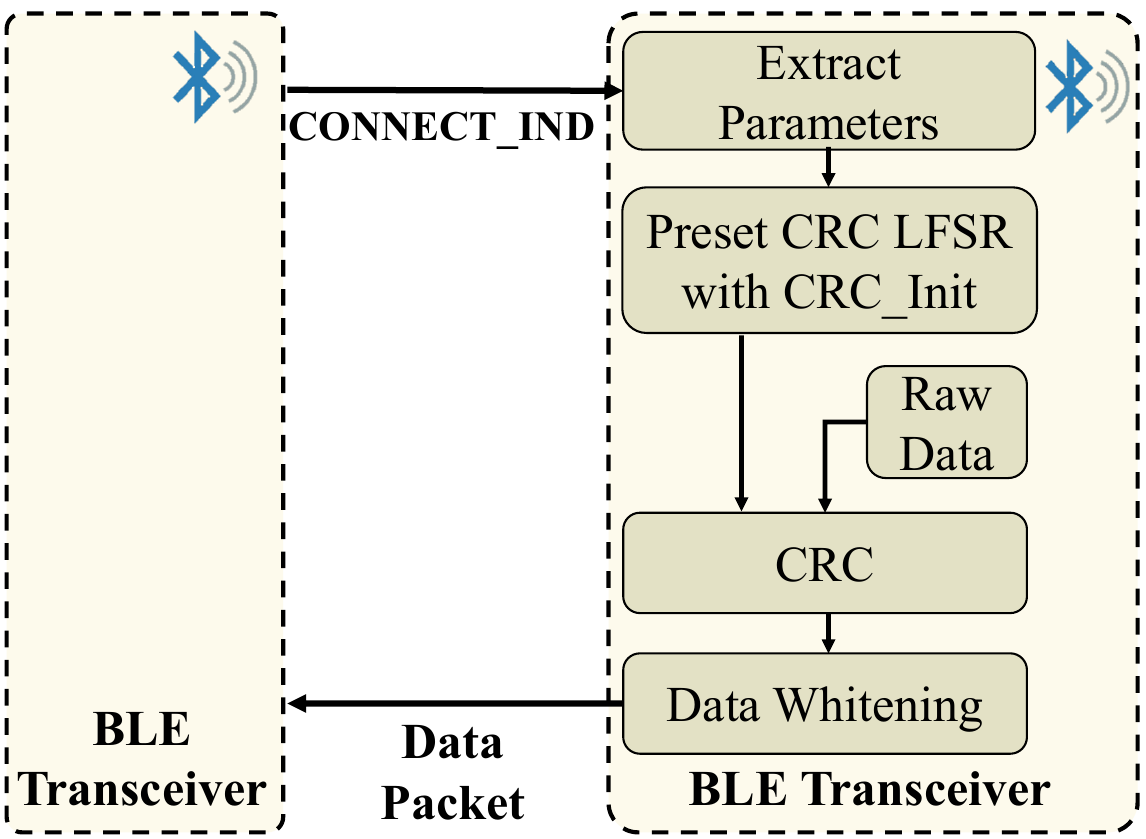}
          }
     \subfigure[BLE Connection on PassiveBLE.]{
			\includegraphics[width=0.48\linewidth]{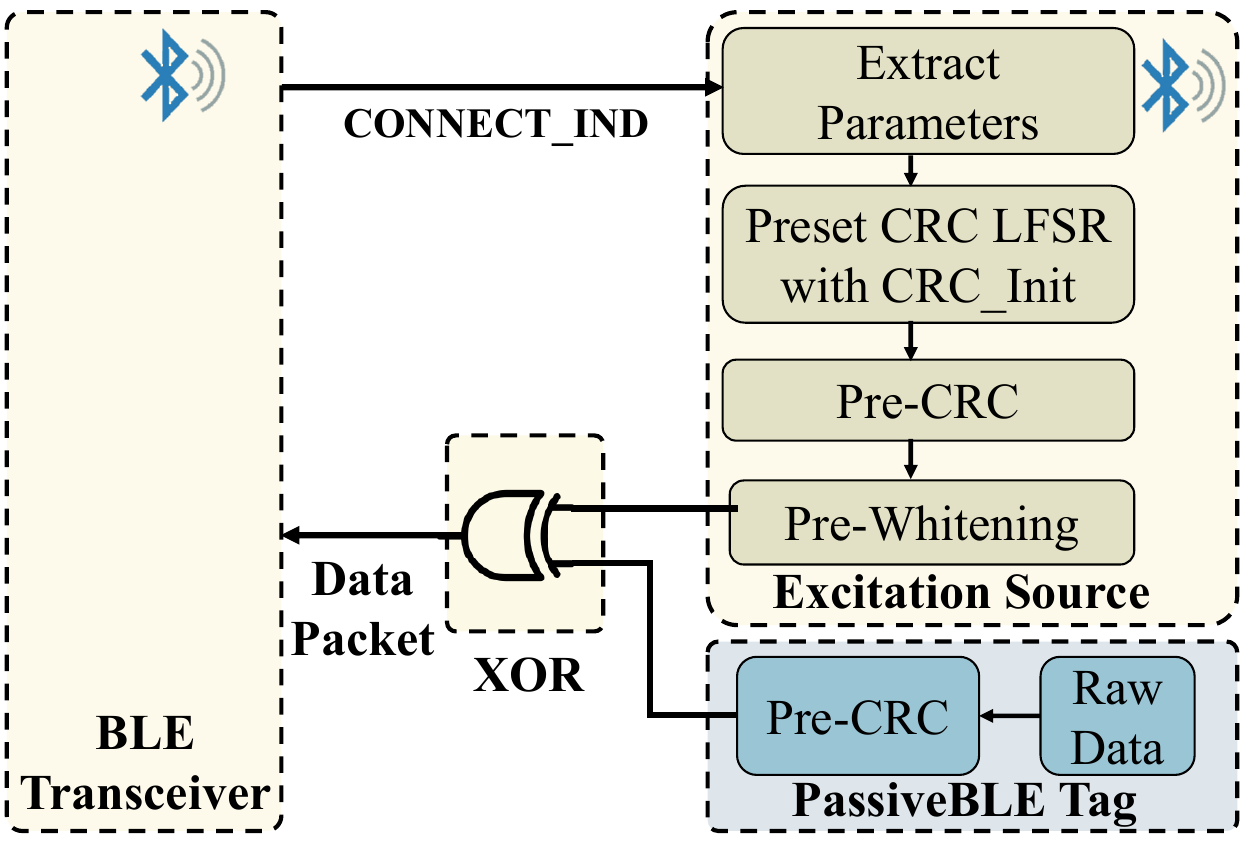}
          }
    \caption{Architecture of active BLE and PassiveBLE.
    }
    \label{fig:Architecture} 
    \end{minipage}
\end{figure*}

To overcome the second challenge, we propose a distributed coding scheme that directly uses the RF signals for logic computing. We observe that the standard BLE systems use a series of operations including exclusive OR (XOR) logical operations to combine the dynamic parameters with the data for encoding.
Particularly, we offload part of the channel coding operations that require dynamic parameters, which can only be acquired by fully decoding incoming packets, to the excitation source. To this end, we reformulate the operations according to the distributive and associative laws, in order to separate operations involving dynamic parameters and other operations as two inputs of the XOR operation. Specifically, the excitation source generates pre-modulated carriers with these parameters, and then the tags use the ``XOR'' operation to generate BLE data packets via scattering the pre-modulated carriers. At a high level, \sysname’s design incorporates an RF computing mechanism that operates RF signals' property for computing purposes. In this way, the backscatter tags do not need to decode the incoming BLE packets to acquire parameters for channel coding.  

To solve the last challenge, we design a connection scheduler to offload the tasks of negotiation and connection management from tags to the excitation source.
To relieve tags from on-board storage and process burdens, we modify the firmware of a commodity BLE transceiver as an excitation source to establish, maintain and terminate connections between tags and unmodified BLE devices. To further reduce the energy cost of tags from frequency hopping, the excitation source generates frequency-hopped carriers to achieve the frequency hopping spread spectrum (FHSS). 
Thus, resource-limited tags can conserve more energy for synchronization and data transmission tasks while retaining high-performance connections with commodity BLE devices.

With all the challenges addressed, we implement the \sysnamenospace, as illustrated in \fig~\ref{fig:setup}. 
The \sysname tag achieves onboard BLE modulation with 491~$\mu$W energy consumption using off-the-shelf components. 
We also design an ASIC sketch using Synopsis Design Compiler~\cite{Synopsis} with a power consumption of about 9.9~$\mu$W.
Then, we conduct extensive experiments to evaluate the performance of \sysname under both line-of-sight~(LoS) and non-line-of-sight~(NLoS) conditions. The results demonstrate that \sysname achieves a success rate of over 99.9\% in establishing commodity BLE connections. \sysname achieves a goodput of up to 532~kbps in \textit{LE 1M PHY} mode, which is about 63.3$\times$ higher than those achieved by the state-of-the-art commodity-level BLE backscatter system~\cite{zhang2021commodity}.
\sysname also achieves a goodput of up to 974~kbps in LoS conditions and up to 966~kbps in NLoS conditions in \textit{LE 2M PHY} mode.  

To the best of our knowledge, \sysname is the first attempt to generate standard BLE packets using backscatter to enable fully compatible BLE connections. Main contributions are summarized as follows.
\begin{itemize}
    \item We provide a low-power coherent-detection-based synchronization circuit for backscatter systems, which provides higher accuracy and lower delay than existing envelope-detection-based approaches. We achieve a sensitivity of up to -33~dBm via off-the-shelf components, with a synchronization accuracy of about 93~ns.
    \item We provide a distributed coding scheme to construct an XOR operation on RF signals, which allows backscatter tags to generate standard BLE data packets without knowing dynamic parameters.
    \item We provide a BLE connection scheduler to manage standard BLE connections between tags and commodity BLE devices. It allows backscatter tags to establish and maintain BLE connections with BLE devices such as smartphones without any modification.
\end{itemize}
\balance

\section{Observation} \label{Observation}

Upon delving into the BLE communication schemes, we find the following two observations to realize BLE connections on backscatter tags. 

First, we observe that the structure of BLE data packets provides an opportunity for the BLE excitation source to wake up the backscatter tags and generate BLE data packets with only software modifications. \fig~\ref{fig:Observation} illustrates the packet structure of an LE 1M data packet. The LE 2M data packet follows a similar structure but features a 2-byte preamble. 
The preamble and header of the BLE packet can be reused to wake up and synchronize the backscatter tag without modification. With software-level modifications, the Access Address can be used to select a specific backscatter tag. The variable-length payload (0-251 Bytes) can be used to generate a shorter BLE data packet with a variable-length payload (0-241 Bytes). Therefore, based on this observation, the activation of backscatter tags as well as BLE data packet generation, can be completed in a short period, which makes the BLE connection on backscatter tags possible.

Second, we observe that the standard BLE systems use XOR logical operations to mix the dynamic parameters with the data for channel coding, e.g., CRC generation and whitening, before modulating the data on an RF carrier~\cite{BLECRCInit}. 
Thus, we can decompose the encoding process into two parts: one on the excitation source to perform pre-modulation and the other on the backscatter tag to pre-process the data to be sent. Then, we can combine these two parts with an external XOR operation to generate commodity-level BLE data packets. Meanwhile, we observe that the toggling of the wireless signals' phase can be used to construct an XOR operation during the scattering process. Based on these observations, we intend to develop a distributive encoding scheme to generate standard BLE data packets on backscatter tags. 
The comparison with conventional BLE data packet generation is illustrated in \fig\ref{fig:Architecture}. We utilize the BLE excitation source to store the dynamic parameters and use these parameters to generate pre-modulated carriers for the backscatter tag. The tag combines the data with the pre-modulated carriers using the XOR operation to generate standard data packets.
\balance

\section{Design} \label{Design}	
\subsection{Design Overview}

\begin{figure}
    \centering
    \setlength{\abovecaptionskip}{0.cm}
    \setlength{\belowcaptionskip}{0.cm}
    \includegraphics[width=\linewidth]{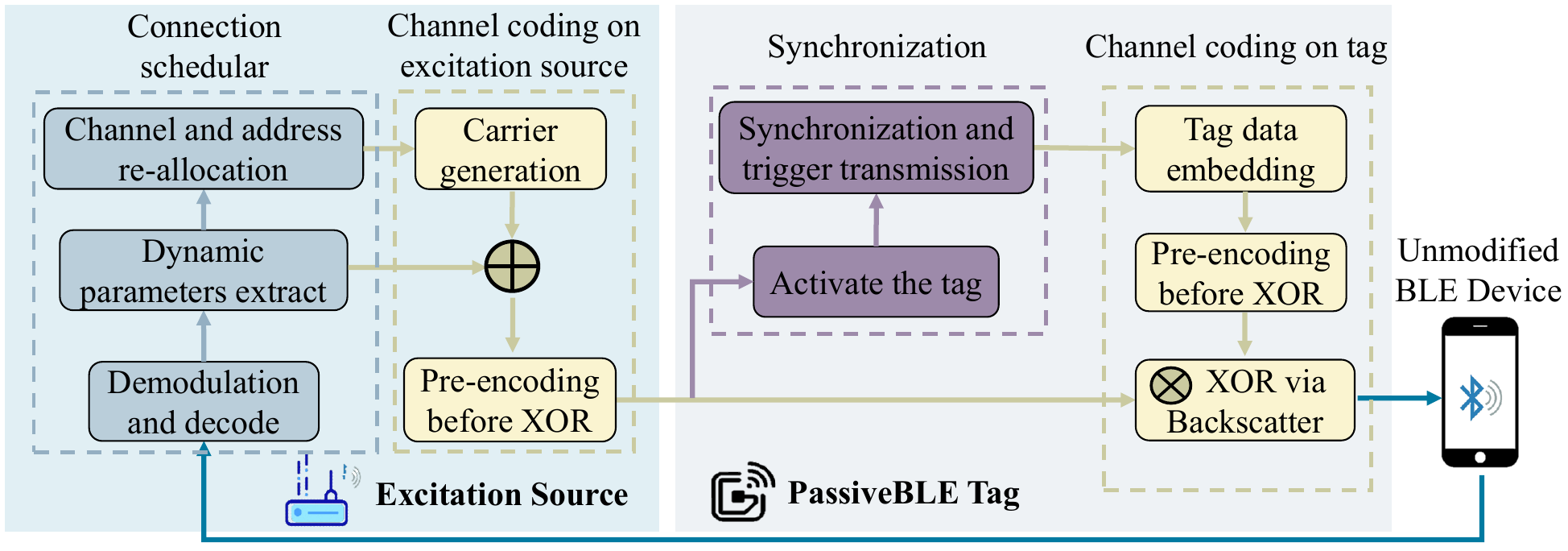}
    \caption{Overview of PassiveBLE system.}
    \label{fig:system-overview}
\end{figure}

As shown in \fig~\ref{fig:system-overview}, the \sysname system consists of tags, an excitation source, and unmodified BLE devices. The excitation source is a commodity BLE transceiver that emits pre-modulated BLE data packets as carriers for backscatter tags. The \sysname tag XOR data on the pre-modulated BLE packets to generate standard BLE packets with data payload. Then, the unmodified BLE device can directly decode the standard BLE packet to extract data from the tag. The excitation source can help one or multiple tags establish, maintain, and terminate standard BLE connections between tags and unmodified BLE devices. 
\sysname employs the following hardware and software key designs to achieve BLE connections between tags and commodity BLE devices.

\textbf{Synchronization.} The synchronization circuit is implemented on the \sysname tag. We propose a high-accuracy, low-latency synchronization circuit to enable $\mu$s-level latency in wake-up and synchronization of a selected \sysname tag and ns-level accuracy to achieve symbol-to-symbol alignment between the excitation source and the tag. This design lays the basis to support the symbol-aligned XOR operation for encoding and quickly responding to commodity BLE devices during the connection.

\textbf{Channel coding.} Our key insight is directly using an RF computing mechanism to achieve lightweight channel coding. The channel coding scheme is implemented on both the \sysname tag and the excitation source. It reduces the burden of encoding and processing operations on tags with dynamic parameters to achieve commodity-compatible high-rate encoding with lower power consumption. This design is the key to generating BLE data packets on resource-limited backscatter tags and achieving physical-layer compatibility with commodity BLE devices.

\textbf{Compatible connection.} The connection scheduler is implemented on the excitation source. It employs the excitation source to conduct all the required control and interactions, e.g., FHSS, establish, maintain and terminate the BLE connections between the backscatter tags and unmodified commodity BLE devices.  

\begin{figure}[t]
    \centering
    \setlength{\abovecaptionskip}{0.cm}
    \setlength{\belowcaptionskip}{0.cm}
    \includegraphics[width=\linewidth]{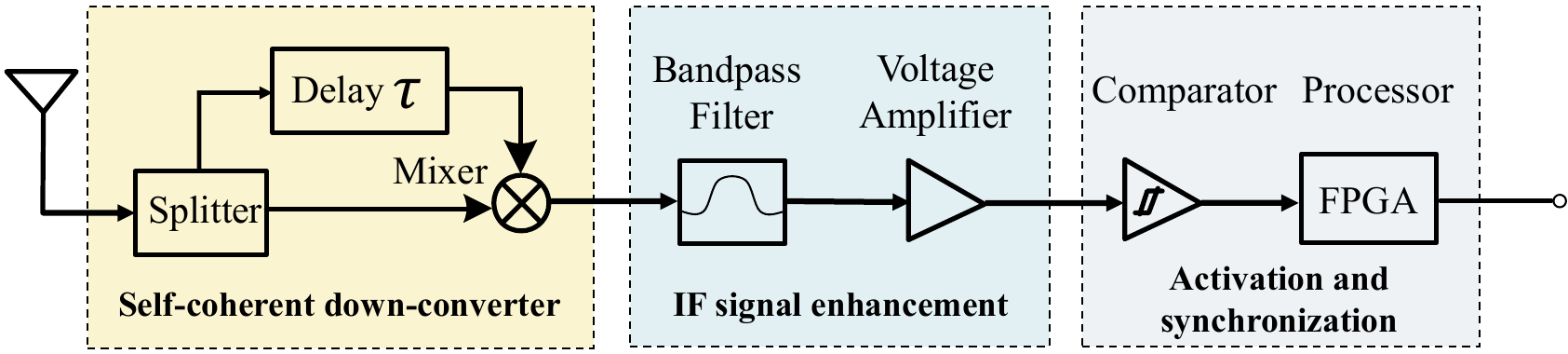}
    \caption{The high-level synchronization circuit schematic of \sysnamenospace.}
    \label{fig:DownlinkCircuit}
\end{figure}

\vspace{-0.3cm}
\subsection{Synchronization} \label{sec:Sync}
The high-level synchronization circuit is illustrated in \fig~\ref{fig:DownlinkCircuit}. 
It consists of three major components including (i) a self-coherent down-converter to extract the frequency difference of neighboring BLE symbols, (ii) an IF signal enhancement module that consists of a filter and an amplifier circuit to strengthen the extracted difference, (iii) an activation and synchronization module which consists of a voltage comparator and logical correlation circuits in FPGA to detect the preamble of BLE packets for synchronization.

\begin{figure*}[t]
        \hfill
         \begin{minipage}{0.44\linewidth}  
                \centering
                \setlength{\abovecaptionskip}{0.cm}
                \setlength{\belowcaptionskip}{0.cm}
                \includegraphics[width=\linewidth]{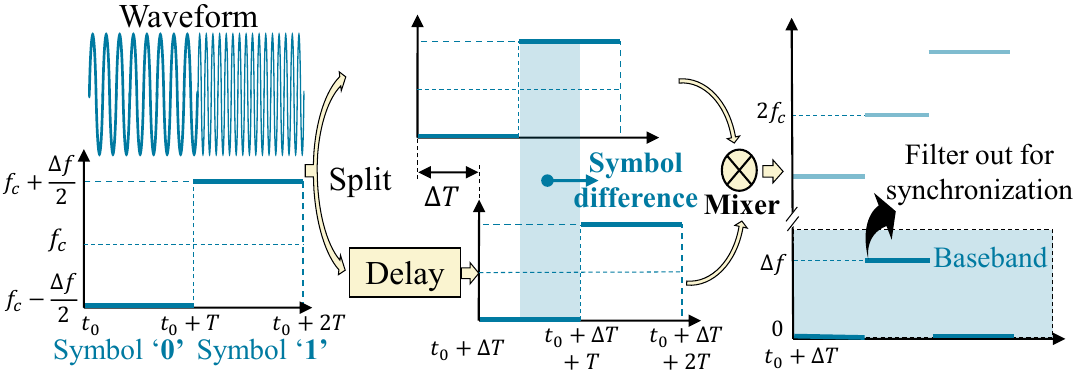}
                \caption{The illustration of the BLE symbol \\ difference extraction.}
                \label{fig:waveformExample}
        \end{minipage}
        \begin{minipage}{0.26\linewidth}  
                \centering
                \setlength{\abovecaptionskip}{0.cm}
                \setlength{\belowcaptionskip}{0.cm}
                \includegraphics[width=\linewidth]{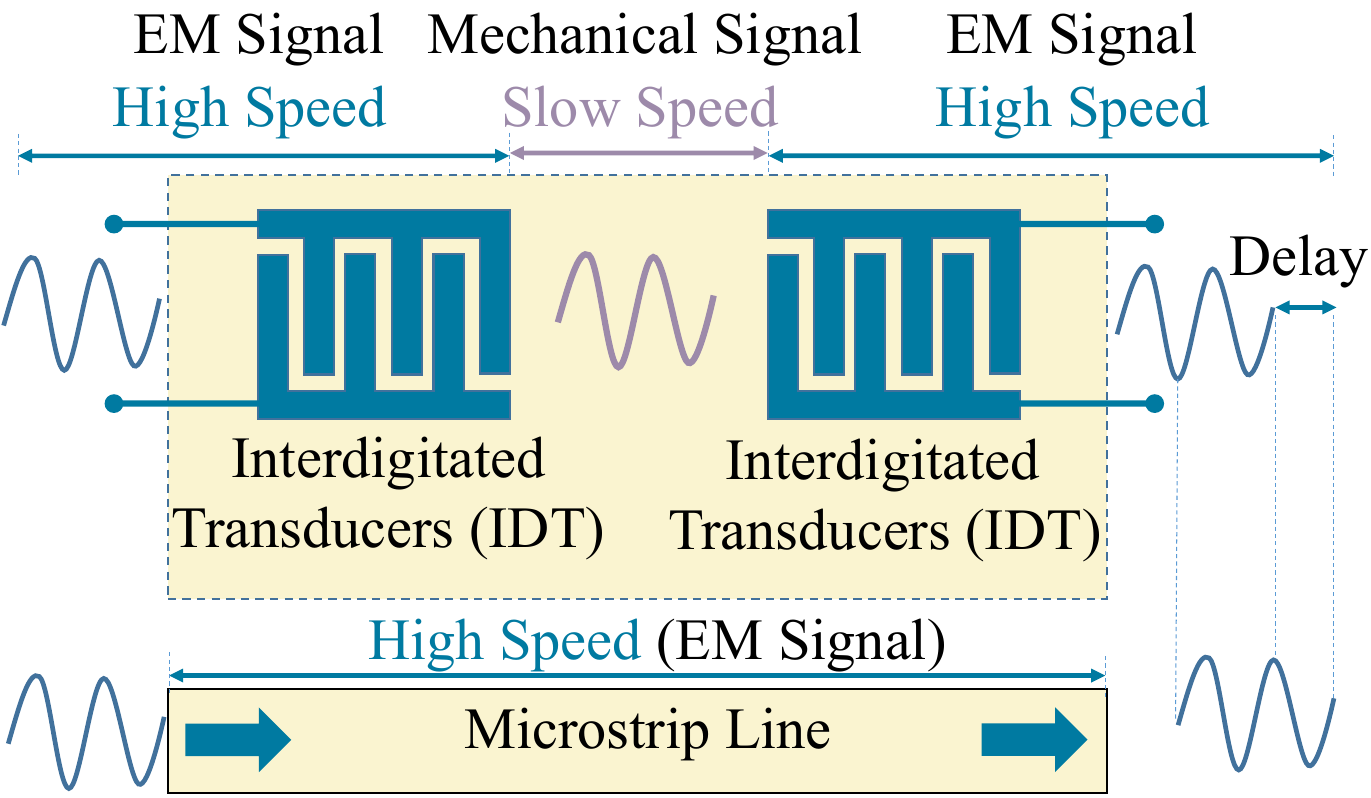}
                \caption{The signal delay \newline effect of SAW filter.}
                \label{fig:SAW_Delay}
        \end{minipage}
        \begin{minipage}{0.29\linewidth}  
                \centering
                \setlength{\abovecaptionskip}{0.cm}
                \setlength{\belowcaptionskip}{0.cm}
                \includegraphics[width=\linewidth]{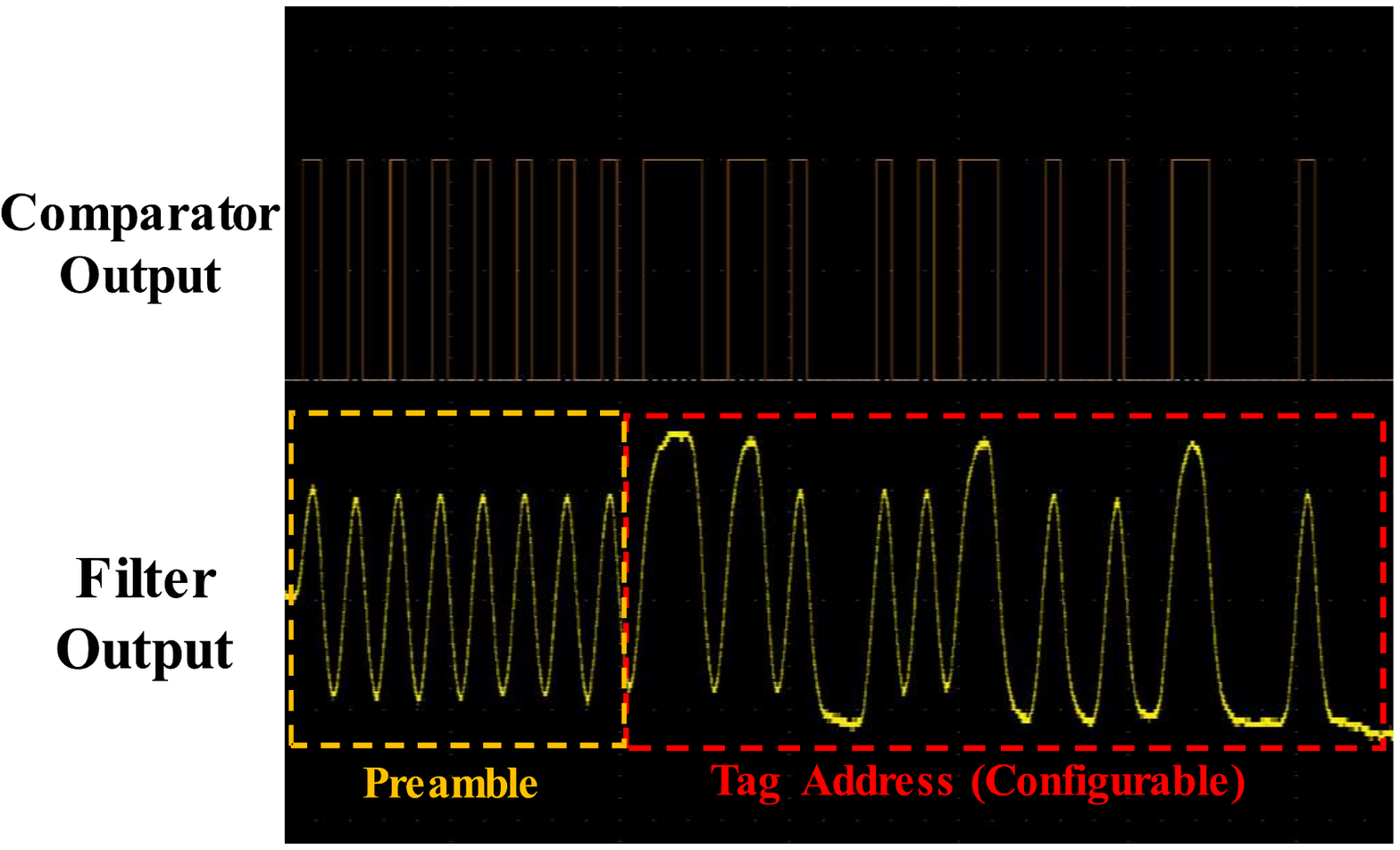}
                \caption{Output showcase.}
                \label{fig:Preamble_Example}
        \end{minipage}
\end{figure*}

\textbf{Self-coherent down-converter.}
BLE signals are modulated with Gaussian Frequency Shift Keying (GFSK), which contains two types of symbols at different frequencies. \fig~\ref{fig:waveformExample} shows a piece of BLE signal at center frequency $f_c$. The symbol ``1'' is represented with electromagnetic waves at frequency $f_1=f_c+\Delta f/2$ and ``0'' at frequency $f_2=f_c-\Delta f/2$.
Upon reception of the signal at the down-converter circuit, the signal splits into two paths and one of the paths will add a $\Delta T$ time delay. Then we mix the delayed signal and the non-delayed signal. If neighboring symbols differ, it will produce an IF signal with a frequency $\Delta f=|f_1-f_2|$, which can be used for synchronization after we filter it out.

\textbf{Delay RF signals.} Delaying the RF signals is challenging in a size-limited backscatter tag. Existing backscatter designs~\cite{zhao2019ofdma} usually use microstrip lines to delay signals. A 64~mm microstrip line can only delay the signal for 1.26~ns~\cite{okubo2024integrated}, which is insufficient to delay BLE symbols ($\geq500$~ns) on a size-limited backscatter tag.
To cope with this issue, we use SAW filter, which is commonly used to attenuate unwanted signals with different frequencies in backscatter systems~\cite{guo2022saiyan, jiang2023dances} or against metallic environments and interference in RFID systems~\cite{suresh2020comparative}, to delay signals.
The principle of signal delay in a SAW filter is illustrated in \fig~\ref{fig:SAW_Delay}. One of the split signals passes through a short microstrip line and the other passes through a SAW filter. The signal passing through the microstrip line incurs negligible delay
due to the high speed of EM signals (close to the lightspeed in vacuum~\cite{andreyev2016temperature}). The signal that passes through the SAW filter is converted to mechanical signals by an interdigitated transducer (IDT), and then propagates as mechanical signals in the SAW from one side to the other. After reaching the other side, the mechanical signals will be converted back to EM signals. As the mechanical signals are much slower (about 5.5~km/s~\cite{naumenko2019linbo}) than the EM signals, a small SAW filter will create a significant delay difference between these two signals. 
After testing various off-the-shelf SAW filters, we choose model EPCOS B8328~\cite{SAW_Filter}. Three cascaded small SAW filters ($1.4 \times 1.1$~mm$^2$) can provide about 46~ns time delay with about 3.3~dB insertion loss (measured by a Keysight MXR608A oscilloscope~\cite{Oscilloscope}). A 3.3~dB amplitude difference between mixed signals does not directly alter the output frequency accuracy, as frequency is determined by phase relationships. Then we build a single diode mixer to combine these two signals. It will output the IF signal with a frequency $\Delta f$, along with the linear combinations of original frequencies $f_1$ and $f_2$, including $f_1 + f_2$, $2f_1$, $2f_2$, etc.


\begin{figure*}[t]
        \hfill
        \begin{minipage}{0.41\linewidth}  
                \raggedleft
    			\includegraphics[width=\linewidth]{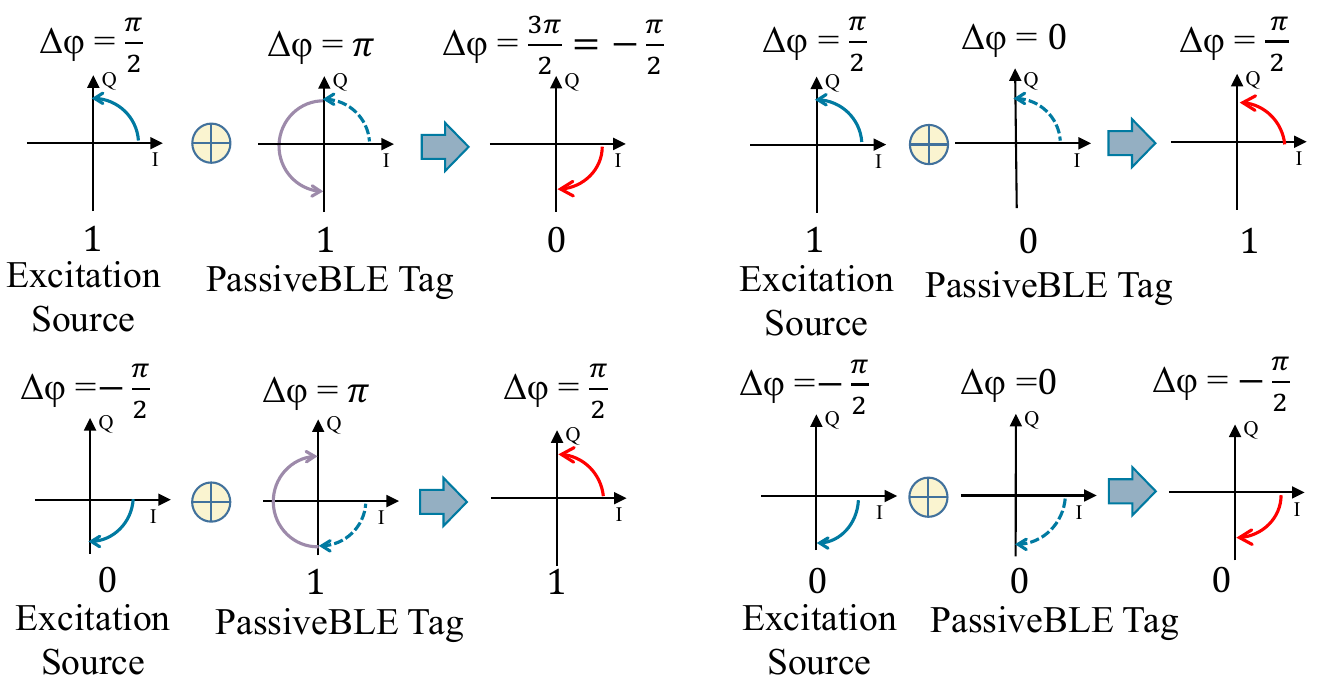}
        \caption{XOR operation via backscatter.}
        \label{fig:XorAir}
        \end{minipage}
        \begin{minipage}{0.58\linewidth}  
                \includegraphics[width=\linewidth]{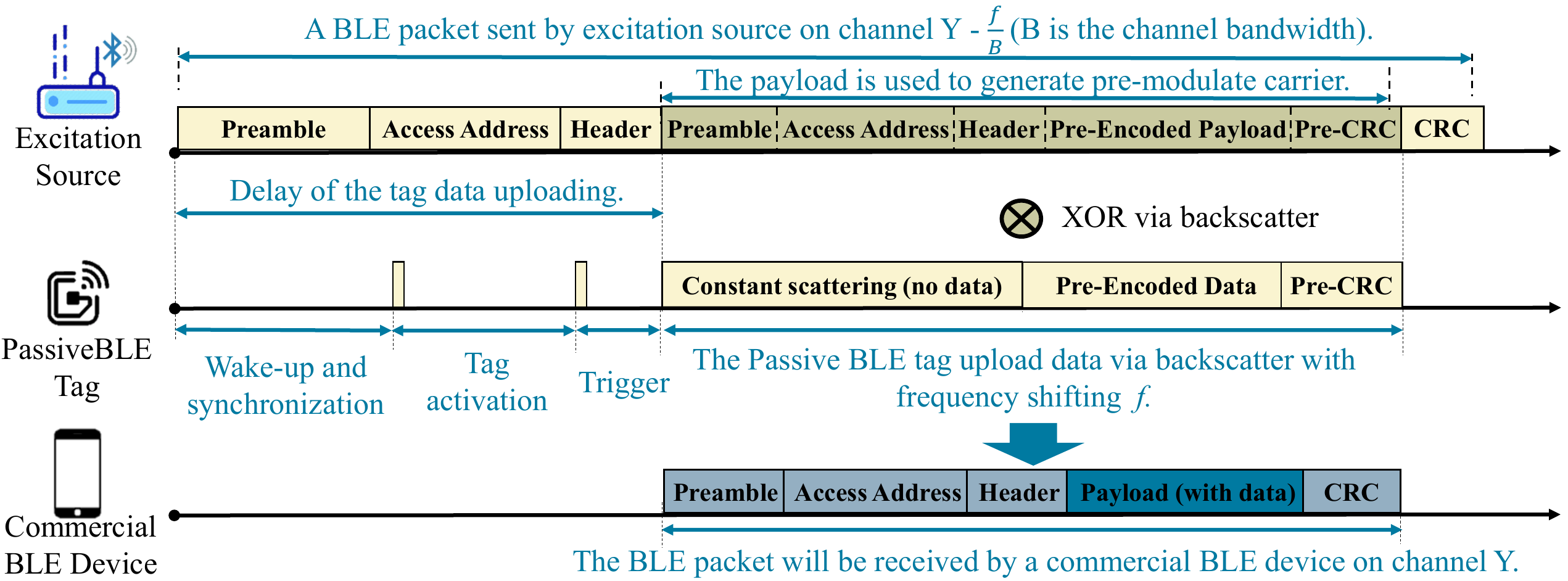}
                \caption{BLE packet generation in a \sysname system.}
                \label{fig:packetGen}
        \end{minipage}
\end{figure*}

\textbf{IF signal enhancement.} To improve the performance of synchronization, we strengthen the IF signals at frequency $\Delta f$ by filtering out unwanted signal and amplifying the IF signals. First, we design a low-pass filter with 3~dB bandwidth to 10~MHz to remove unwanted original signals, and their linear combinations. Then we implement a high-impedance voltage amplification circuit, which provides about 30~dB gain with $\mu$W-level power to amplify IF signals.

\textbf{Activation and synchronization.} The ability to extract the difference between symbols provides an opportunity to reuse the structure of BLE packets for tag activation and synchronization. Therefore, the preamble of the BLE packets, which is used for synchronization in commodity BLE devices, can also be used to wake up and synchronize the tags. The Access Address of the BLE packets, which is used to identify different BLE connections, is leveraged to identify different tags and support multiple BLE connections with tags in \sysnamenospace. Meanwhile, as the Access Address is configurable, we can also re-allocate the Access Address with continuous ``00'' to stand for a bit ``0'' and ``01'' stands for a bit ``1'' to improve the sensitivity. As a tradeoff, the maximum number of connections that \sysname can support will be reduced. For example, the 4 Bytes Access Address can be reused to allocate to $2^{32/n}$ tags if we use $n$ different BLE symbols to stand for a bit ``1''. As shown in \fig~\ref{fig:Preamble_Example}, we present a circuit output captured using a Tektronix MSO64B oscilloscope, clearly demonstrating that the preamble and tag address from an LE 2M packet can be successfully extracted. 

\vspace{-0.3cm}
\subsection{XOR Operations on RF Signals}\label{Design:XOR}
To overcome the challenge that tags do not have LO and insufficient processing capability to demodulate and decode BLE packets to extract the dynamic parameters, we observe in \S~\ref{Observation} that the standard BLE channel coding process can be achieved by XOR-ing dynamic parameters at the excitation source with data on a tag. 
Thus, we intend to construct XOR operations directly on RF signals, which is called RF Computing~\cite{na2023leggiero}, to build standard BLE packets with backscatter.

On one hand, commodity BLE transceivers decode the BLE symbols by estimating the phase accumulation $\Delta \phi$ during the symbols, where $\Delta \phi = \frac{\pi}{2}$ represents bit "1" and $\Delta \phi = - \frac{\pi}{2}$ represents bit "0". 
On the other hand, a tag shifts the BLE signals with a square wave to another channel and modulates bits by changing the phase of the square wave~\cite{zhang2021commodity}.
Thus, by taking the phase accumulations caused by the excitation source and the phase change caused by the tag as input, and taking the phase demodulated by the unmodified BLE device as output, an XOR operation is constructed.

\fig\ref{fig:XorAir} illustrates how XOR operations are constructed in \sysnamenospace. 
When the excitation source provides bit "1" by generating a symbol with $\frac{\pi}{2}$ phase accumulation and the \sysname tag provides bit "0" with a constant frequency shift with phase accumulation $0$, the commodity BLE receiver will decode a bit "1" since the overall phase accumulation is $\frac{\pi}{2}$.  
The other three input cases are also consistent with the logic law of XOR operations. Consequently, the results in the receiver only contain two states and meet the logical law of XOR operations.

\begin{figure*}[t]
        \hfill
        \begin{minipage}{0.29\linewidth}  
                \raggedleft
                \setlength{\abovecaptionskip}{0.1cm}
        			\includegraphics[width=\linewidth]{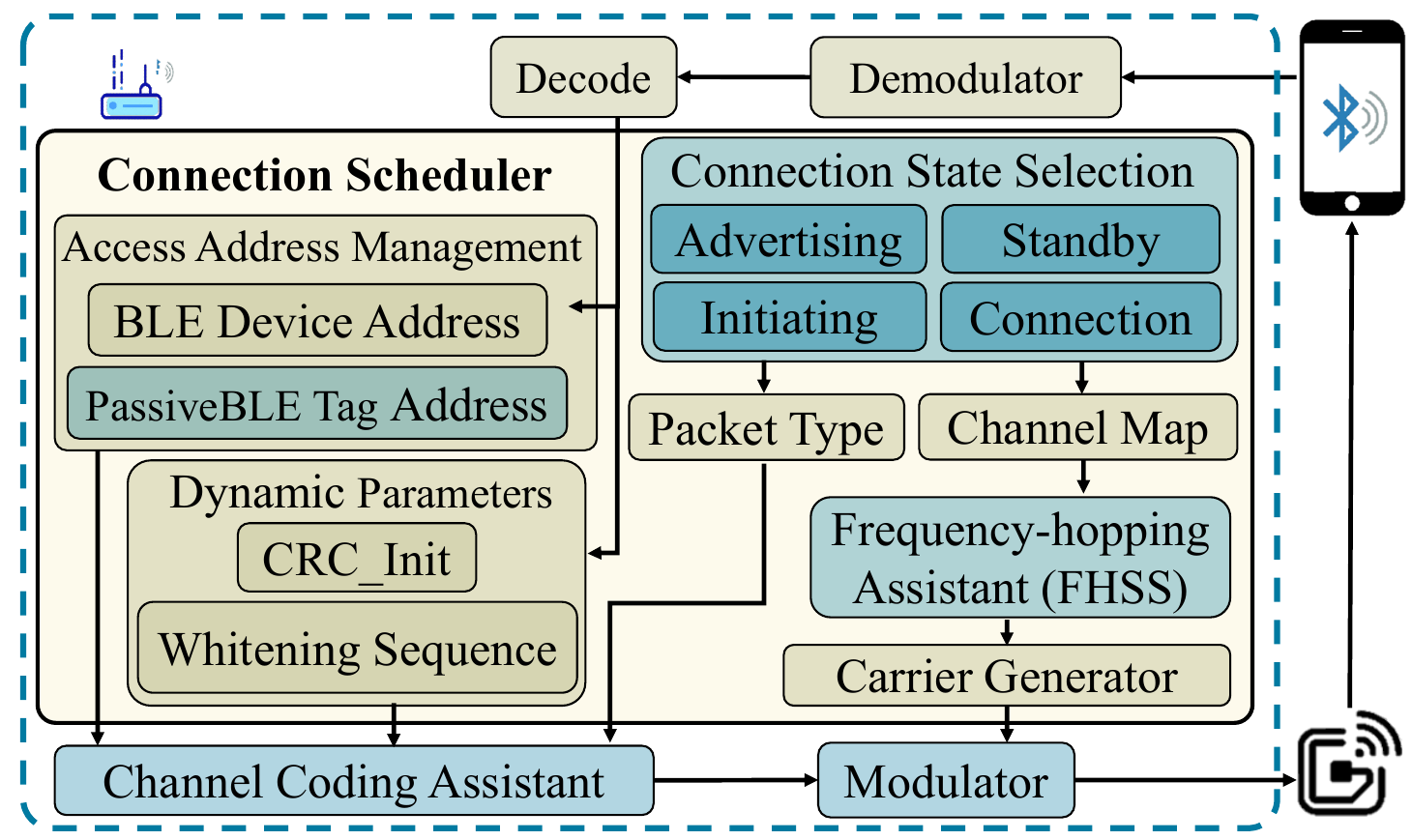}
            \caption{Connection \newline scheduler structure.}
            \label{fig:connectionScheduler}
        \end{minipage}
        \begin{minipage}{0.50\linewidth}  
                \raggedleft
                \setlength{\abovecaptionskip}{0.1cm}
    			\includegraphics[width=\linewidth]{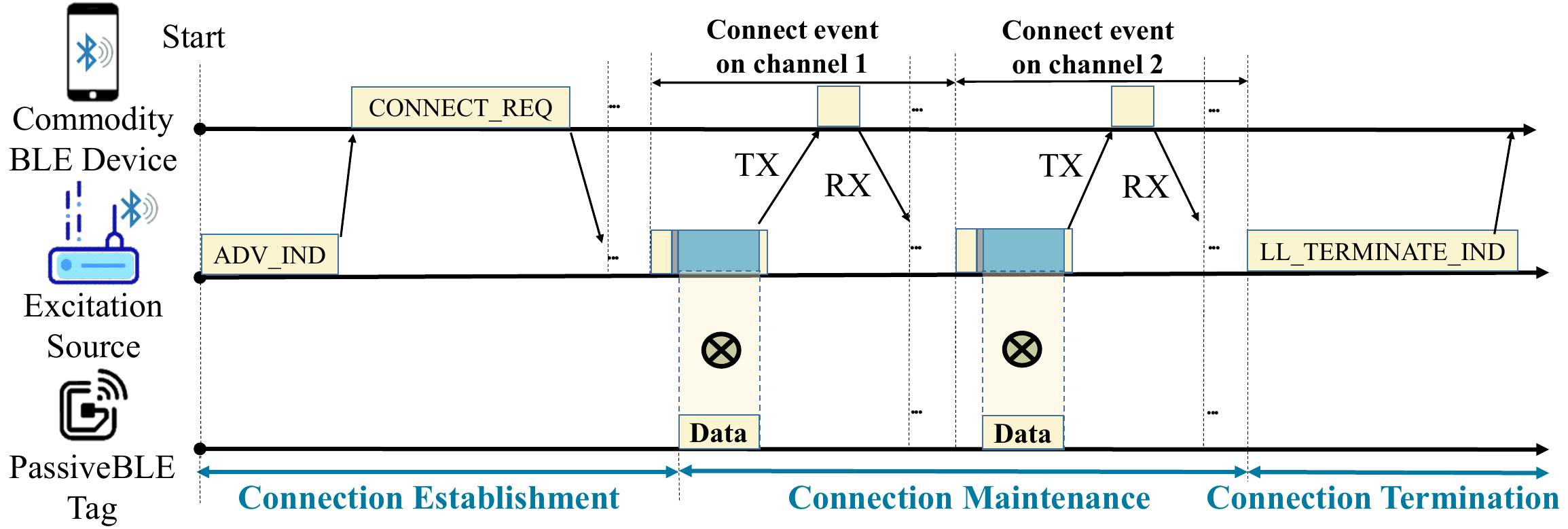}
        \caption{A connection lifetime example for a \sysname tag and a commodity BLE device.}
              \label{fig:connect_event_example}
        \end{minipage}
        \begin{minipage}{0.19\linewidth}  
                \raggedleft
                \setlength{\abovecaptionskip}{0.1cm}
    			\includegraphics[width=\linewidth]{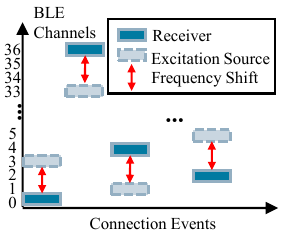}
        \caption{FHSS on tag.}
        \label{fig:FHSS}
        \end{minipage}
\end{figure*}
\vspace{-0.3cm}
\subsection{BLE Packets Generation with XOR Operations}
\fig~\ref{fig:packetGen} explains how the XOR logical operation can be used to achieve BLE-compatible CRC generation and data whitening. After connection establishment, the unmodified BLE device listens channel $Y$ for data packet reception. Then, the excitation source will send a BLE packet to activate the tag and provide a carrier for the tag to transmit data at channel $Y-f_{s}/B$, where $f_{s}$ is the frequency shift of the tag and $B=2$~MHz is the bandwidth of a BLE channel. The Preamble, Access Address, and Header of the packet will be used to activate the tag. We reallocate the first several bytes (seven bytes in LE 1M PHY or eight bytes in LE 2M PHY) of the original payload as the Preamble, Access Address, and Header of a new BLE packet sent by tags. These bytes are encoded by the excitation source and frequency-shifted by the tag to channel $Y$ without phase changes. The rest of the payload is used by the tag as the new payload. Then, the tag encodes the message with CRC generation and data whitening with XOR operations, which is explained below.

 \textbf{CRC generation.}
After a BLE connection is established, dynamic parameters including \textit{CRC\_Init} and \textit{Channel Index}, are determined. \textit{CRC\_Init} is the initial value of the CRC generator, and \textit{Channel Index} is the index of the current channel of the BLE connection. When the BLE connection hops into a new channel, \textit{Channel Index} changes.

In standard BLE systems, a transceiver obtains the dynamic parameter \textit{CRC\_Init} by decoding a received packet and then uses \textit{CRC\_Init} to generate a CRC checksum, which is used in the following packet the transceiver transmits. However, backscatter tags cannot decode the BLE packets due to the lack of LO and insufficient processing capability. 

To address this issue, \sysname tags are designed to generate the correct CRC checksum without knowing the \textit{CRC\_Init}. Given the message to be sent by the tag $m(x)$ and the initial state \textit{CRC\_Init} of the CRC generator, which is a 24-bit LFSR circuit that can be expressed as $g(x)=x^{24}+x^{10}+x^9+x^6+x^4+x^3+x+1$, the CRC checksum \textit{CRC\_Value} according to the BLE specification can be computed as~\cite{BLECRCInit}
\begin{equation}
CRC\_Value = (m(x) \oplus CRC\_Init)~\%~g(x) \label{eq:1}
\end{equation}
where $\oplus$ is the XOR operation and \textit{\%} is the modulo operation. Note that the modulo operation is implemented in the circuit using a series of modulo 2 subtractions, which satisfies the distributive law. 
Therefore, Eq.~(\ref{eq:1}) can be expressed as
\begin{equation}\label{eq:2}
  CRC\_Value = (m(x)~\%~g(x))\oplus (CRC\_Init~\%~g(x))  
\end{equation}
Thus, the tag modulates $(m(x)~\%~g(x))$ and the excitation source modulates $(CRC\_Init~\%~g(x))$. Then, the tag and the excitation source generate $CRC\_Value$ by performing the XOR operation. By this means, without knowing the $CRC\_Init$ bits, \sysname tag can also generate correct $CRC\_Value$ over the air and create standard BLE packets with the assistance of the excitation source.

\textbf{Data whitening.}
Data whitening is used in BLE to enhance the stability of wireless links by avoiding long sequences of zeros or ones. Before transmitting packets, both the header and the payload undergo scrambling with a data whitening sequence to randomize bit sequences, reduce highly redundant patterns, and minimize DC bias. In BLE systems, the BLE signal $S_{W}(x)$ after whitening and modulation can be expressed as 
\begin{equation}
  S_{W}(t) = (m(x) \oplus w(x)) \times e^{2\pi f_c t}
\end{equation}
where $m(x)$ is the message to be sent. $w(x)$ represents the whitening sequence and is derived from the current \textit{Channel Index}~\cite{BLECRCInit}. $e^{2\pi f_c t}$ is the carrier signal.
Since XOR operation satisfies the associative law, $ S_{W}(t)$ can also be derived by
\begin{equation}
    S_{W}(t) = m(x) \oplus (w(x) \times e^{2\pi f_c t}) \label{eq:4}
\end{equation}
Thus, the excitation source modulates $(w(x) \times e^{2\pi f_c t})$. Overall, the pre-modulate process can be briefly concluded with the following equation: 
\begin{equation}
   E_{seq}= CRC\_Init~\%~ g(x)  \oplus w(x)
\end{equation}
where the $CRC\_Init$ \% $g(x)$ is the pre-processed CRC results and $w(x)$ is the whitening sequence for the data whitening process. The configuration and data loading process has a comparable level of complexity to the previous BLE backscatter systems (e.g., InterScatter~\cite{InterScatter}, IBLE~\cite{zhang2021commodity}, etc.), which also load pre-modulated data in the payload part of BLE advertising packets to generate single-tone carriers. The tag only needs to modulate $(m(x)~\%~g(x))$ and perform the XOR operation according to Eq.\eqref{eq:2} and Eq.\eqref{eq:4}.

\vspace{-0.3cm}
\subsection{\sysname Connection Scheduler}

So far we have accomplished BLE data packet construction on a \sysname tag. The final step towards achieving standard-compatible BLE connections is to implement the commands and interactions for connection establishment, maintenance, and termination. To achieve this goal, we design a connection management scheduler on the excitation source to assist \sysname tags to establish, maintain, and terminate connections with unmodified commodity BLE devices. The structure of the scheduler is illustrated in \fig~\ref{fig:connectionScheduler}, where our modifications are highlighted with light blue and the reused BLE processing modules are in yellow. We employ three new components to enable the connection scheduler, including (i) a connection state management module to determine the proper packet to be sent in different connection states, (ii) a frequency-hopping assistant to help the \sysname tags accomplish the FHSS mechanism according to BLE specifications, and (iii) an access address management module to reallocate addresses for multiple \sysname tags. 

\textbf{Connection state management.} The connection state selection module assists \sysname tags during the whole connection lifetime, which consists of four basic states, including advertising, initiating, connection, and standby states. In the advertising state, the scheduler can help tags send and receive advertising packets. In the initiation stage, the scheduler helps the \sysname tag negotiate with the active BLE devices and receive the dynamic parameters for the connections. In the connection state, the scheduler extracts the channel map to assist the frequency hopping during different connection events. In the standby mode, the scheduler helps tags terminate BLE connections. \fig~\ref{fig:connect_event_example} illustrates a connection lifetime example. First, an excitation source sends an advertising packet, which can also sent by tag, to request a BLE connection with a tag. Then, the commodity BLE device answers the request to initiate connection establishment between the tag and the BLE device. Next, the tag XORs the data on the BLE data packets generated by the excitation source to generate standard BLE packets for commodity BLE devices. After several connection events, the excitation source, or the BLE device, can terminate the connection by sending a connection termination control packet. 

\textbf{Frequency-hopping assistant.}
We employ the excitation source to help tags perform frequency hopping by generating carriers at different channels due to the following reasons: (i) FHSS requires the $\textit{Channel Index}$ which cannot be obtained by tags, and (ii) synthesizing multiple frequencies is difficult and power-hungry for an energy-limited \sysname tag. As illustrated in \fig\ref{fig:FHSS}, we use the excitation source to extract the frequency hopping pattern from the \textit{CONNECT\_IND} packet and provide a carrier with a dynamic frequency $f_t-f_s$, where $f_t$ is the desired frequency according to FHSS at time $t$. The \sysname tag modulates signals with a fixed frequency shift $f_s$. Consequently, the \sysname tag only needs to transmit the data upon being triggered without knowing the frequency hopping pattern. 
The frequency hopping operation of the PassiveBLE system can be briefly concluded into the following 3 steps. First, the excitation source takes the channel hopping map and negotiates with commodity BLE devices to determine the channel for connection dynamically. Then, the excitation source generates the pre-modulated packets on a channel that is $f_s$ (frequency shift by the PassiveBLE tag) away from the target channel. Finally, the tag shifts the frequency and generates commodity-compatible packets on the target channel. The synchronization of the tags is based on the preamble of the BLE data packets. Since the frequency hopping is determined by the excitation source, the PassiveBLE tags only need to detect the preamble and access address of the data packets, if the tag detects the correct preamble and access address, it will directly modulate the data on the payload part of the packets. The frequency hopping will be finished with the excitation source hopping in different channels.

\textbf{Access Address management.} 
When a connection is established, each BLE device is allocated a unique Access Address.
To allow \sysname tags to communicate with unmodified commodity BLE devices, The excitation source stores Access Address for each tag and adds it to the header of the \sysname packets.
When there are multiple connections with different \sysname tags, the excitation source manages these connections through a time division manner. 

\begin{figure*}[t]
         \setlength{\abovecaptionskip}{0cm}
        \setlength{\belowcaptionskip}{0cm}
        \begin{minipage}[b]{0.24\linewidth}  
                \raggedleft
				\includegraphics[width=\linewidth]{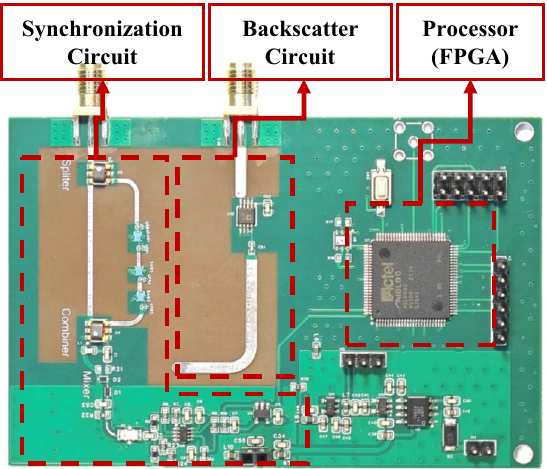}		
                \caption{\sysname tag.} \label{fig:Prototype}
                \label{fig:up_down}
        \end{minipage}
        \hspace{0.3cm}
        \begin{minipage}[b]{0.69\linewidth}  
                \begin{minipage}[t]{0.49\linewidth}
                        \raggedleft
                        \subfigure[Wake-up rate.]{
                                \includegraphics[width=\linewidth]{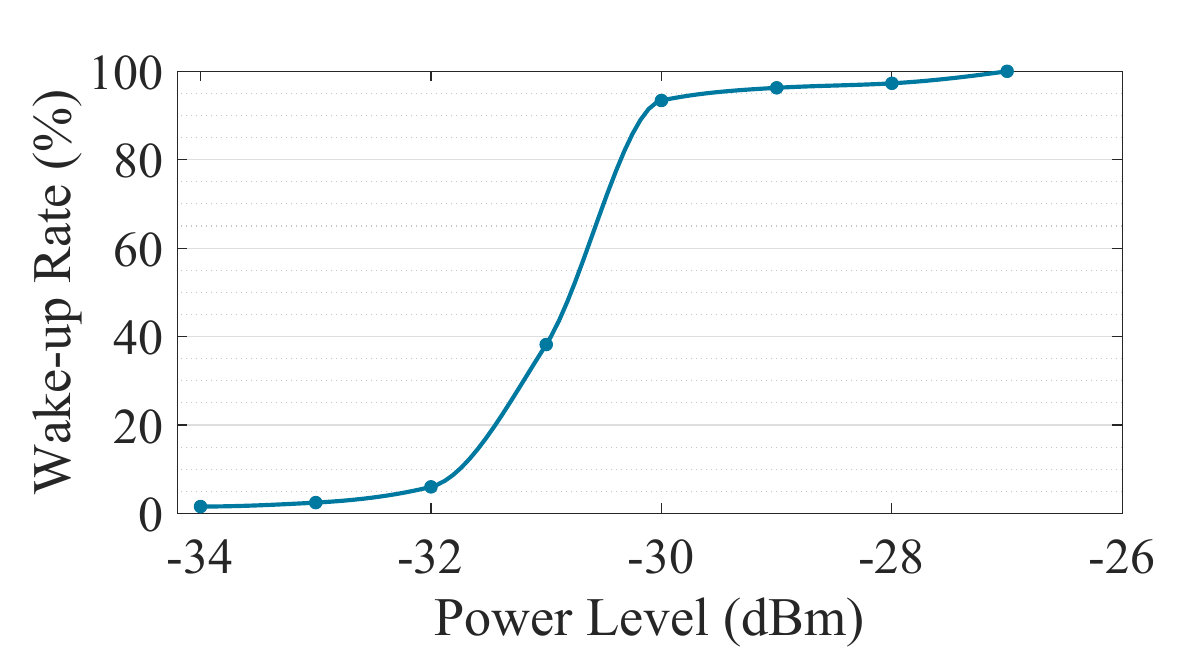}}
                \end{minipage}
                \begin{minipage}[t]{0.49\linewidth}
                        \raggedleft
                        \subfigure[Synchronization jitter.]{
                                \includegraphics[width=\linewidth]{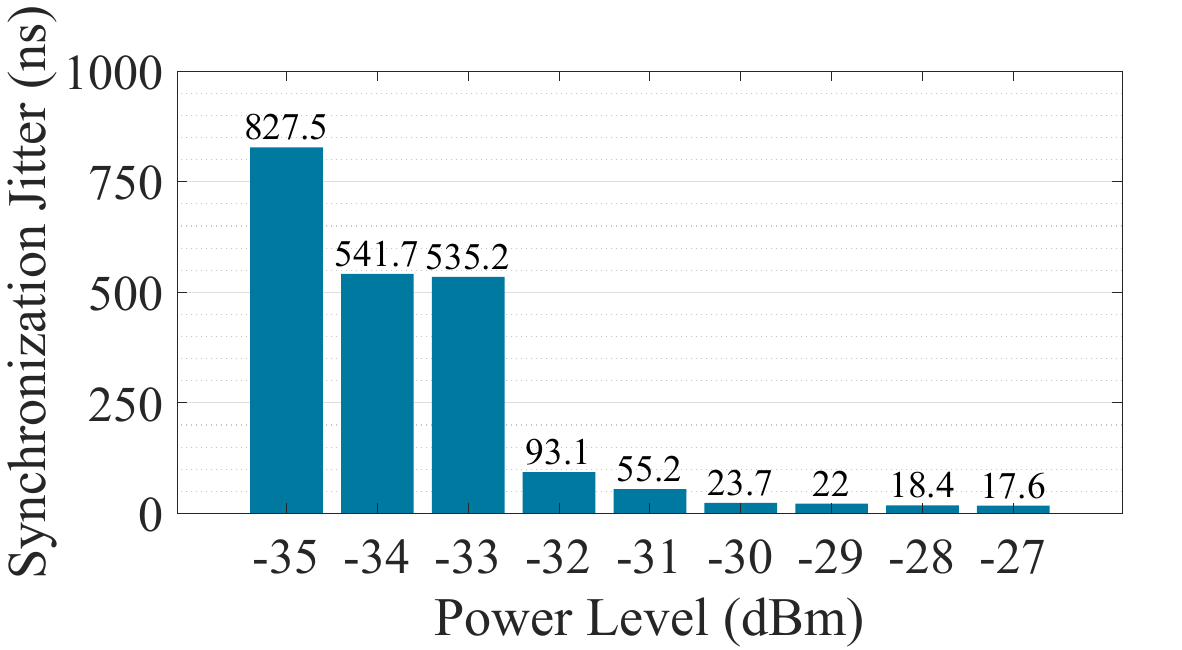}}
                \end{minipage}
                \caption{Wake-up and synchronization performance.}
                \label{fig:Synchronization_accuracy}
        \end{minipage}  
        \vspace{-0.3cm}
\end{figure*}

\section{Implementation}\label{sec-implementation}

\subsection{BLE Transceivers} 
A \sysname system contains a BLE excitation source to generate carriers for \sysname tags. We use a BLE development kit, model Nordic nRF52833 DK~\cite{nRF52833DK} to implement the excitation source.
The excitation source supports both PHY modes defined in BLE, i.e., the \textit{LE\_1M\_PHY} mode that is compatible with all existing BLE devices, and the \textit{LE\_2M\_PHY} mode that is introduced by BLE protocol stack version 5.0. 
The Equivalent isotropic radiated power (EIRP) of the excitation source is set to 20~dBm, which is compatible with the frequency regulation for BLE devices. These operations to configure a BLE transceiver to be an excitation source can be easily achieved with the open API of the Nordic BLE board. A detailed tutorial can be accessed at the product's website~\cite{Nordic52833}.
The BLE devices, such as the receiver, can be any BLE device without firmware or hardware modification, such as unmodified smartphones.

\vspace{-0.3cm}
\subsection{\sysname Tags}

\textbf{COTS implementation.}
As shown in \fig\ref{fig:Prototype}, we implement \sysname prototypes with off-the-shelf components. A \sysname tag consists of three major parts, a synchronization circuit to activate and trigger the tag for communication, an RF switch-based backscatter circuit for uplink communication and an ultra-low-power FPGA (IGLOO AGLN060~\cite{AGLN060}) for baseband processing. 
The synchronization circuit contains the following passive lumped components, including an impedance matching network and a bandpass filter implemented with inductors and capacitors, three cascade SAW filters (	
Qualcomm B39242B8328P810~\cite{SAW_Filter}) to delay the signal, a single-diode mixer implemented with a splitter, a combiner implemented with microstrip lines, and a Schottky diode (Infineon BAT63~\cite{Scotty_Diode}), which can be used to build single-diode mixer with -55~dBm sensitivity~\cite{rostami2021mixiq} in 2.4~GHz frequency band.
We also use two low-power baseband amplifiers ( Infineon LTC6262~\cite{Operational_Amplifier}) and a comparator (ST TS3201~\cite{Comparator}) to convert the signal to binary bits for synchronization. 
The backscatter circuit comprises an RF switch (ADI ADG902~\cite{ADG902}) with microstrip lines.

\textbf{Power Consumption.} All of the active components are powered with a 1.8 V DC supply. There are two modes for the PassiveBLE tag, i.e., uplink mode and standby mode. In the active mode, the tag will upload data via scattering and the synchronization circuit will be disabled. The overall power consumption in this mode is about 491~$\mu$W, including 489~$\mu$W dynamic power consumption of the FPGA and 2 $\mu$W cost by the RF switch. The PassiveBLE works in the standby mode waiting for activation after the transmission. In this mode, overall power consumption is about 1 mW, 
including 35 $\mu$W consumed by the FPGA, 864$\mu$W consumed by the amplifiers, and 131 $\mu$W consumed by the comparator. 
In active mode, the energy consumption of a \sysname tag with both modes enabled is about 1.5~$m$W, 
which is much lower than the existing BLE backscatter system and BLE chips. But current COTS design cannot support low-power standby mode due to the poor static leak current of the off-the-shelf components and the full-duty-cycled synchronization circuit. Engineering efforts from these two aspects will further reduce the power level, and we believe it will perform better than BLE transceivers after being integrated into IC chips with less leakage current.

\begin{table}[h]   
    \setlength{\abovecaptionskip}{0.cm}
    \setlength{\belowcaptionskip}{-0.2cm}
\small
\begin{center}   
\caption{Performance of existing BLE backscatter.}  
\label{Table:ImplementPower} 
\begin{tabular}{|c|c|c|c|}   
    \hline 
     \textbf{Systems} & Power & Goodput & Connection  \\   
    \hline   
    \sysname  &  \makecell{491 $\mu$W \\9.9$\mu$W (IC)} & 974 kbps & $\surd$ \\ 
    \hline
    DanBlue~\cite{jiang2023dances} & 825.4 $\mu$W (IC) & 30~kbps & $\surd$\\
    \hline
    BiBlue~\cite{jiang2023bidirectional} & 240.9 mW & 17~kbps & $\surd$\\
    \hline
    IBLE~\cite{zhang2021commodity} & 2 mW & 8.4~kbps &  ${\times}$\\
    \hline
    Bitalign~\cite{huang2024bitalign} & 1.9 mW & 131.7~kbps &  ${\times}$\\ 
    \hline
    RBLE~\cite{zhang2020reliable} & 37 $\mu$W (IC) & 16.6~kbps &  ${\times}$\\
    \hline
    FreeRider~\cite{zhang2017freerider} & 30 $\mu$W (IC) & 57~kbps &  ${\times}$\\
    \hline
\end{tabular}   
\end{center}   
\vspace{-0.3cm}
\end{table}

\textbf{ASIC design.}
To explore the potential power consumption reduction performance, We also conduct an IC design using Synopsis Design Compiler~\cite{Synopsis} with TSMC 90~nm CMOS process. 
We provide a Verilog description of the digital state machine for baseband processing, BLE packet encoding and frequency shift. Then use the standard cell to simulate the Application Specific Integrated Circuit (ASIC) of \sysname tag.
The static power consumption is 0.9~$\mu$W and the total dynamic power consumption for uplink is 9.9~$\mu$W. We have also simulated the power consumption of the wake-up and synchronization circuits, which consume 190.3~$\mu$W in total, where the amplifier consumes 162.4~$\mu$W and the comparator consumes 27.9~$\mu$W. The power consumption of the synchronization circuit is smaller than  
In Table~\ref{Table:ImplementPower}, we compare both COTS and ASIC designs of \sysname tag with commodity BLE chips and existing BLE backscatter designs, including  
DanBlue~\cite{jiang2023dances} and IBLE~\cite{zhang2021commodity}, etc, to show the potential of \sysname for commodity-level BLE communications.
As shown in Table~\ref{Table:ImplementPower}, we see that a \sysname tag consumes less than 24.8\% power to achieve 7-115$\times$ goodput compared to existing BLE backscatter systems.
Meanwhile, a \sysname tag also consumes 10-100$\times$ less power to achieve 72.8\% throughput of commodity BLE chips~\cite{nRF52833DK}.

\textbf{BLE connection capability.} We also list the measured goodput and connection capability of existing BLE backscatter systems in Table~\ref{Table:ImplementPower}, there are two existing systems, i.e. DanBlue~\cite{jiang2023dances} and Biblue~\cite{jiang2023bidirectional}, can support the BLE connections. But their connections are not commodity-compatible, which needs to pre-allocate static coding parameters for the connections. In standard BLE protocols, as addressed in Section 2 in the manuscript, these parameters are dynamically allocated in commodity BLE devices instead of pre-allocating. Unlike these systems, our system is much more commodity-compatible since it can support the BLE connection with dynamic parameters.

\begin{figure*}[t]
        \begin{minipage}{0.25\linewidth}  
                \centering
                \vspace{-0.03cm}
                \setlength{\abovecaptionskip}{0.cm}
                \setlength{\belowcaptionskip}{0.cm}
                \includegraphics[width=\linewidth]{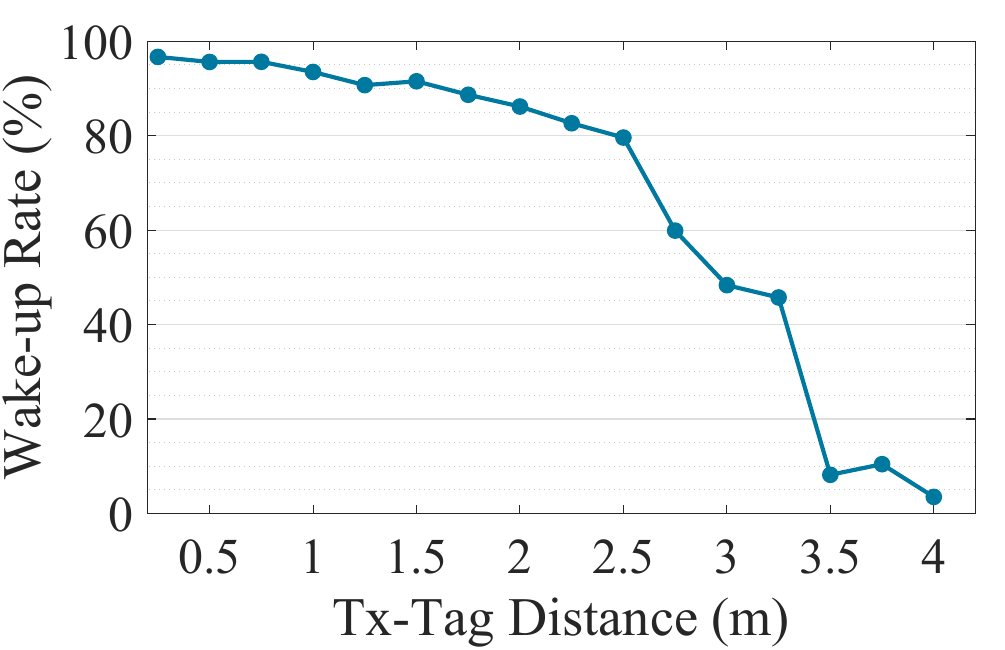}
                \caption{Wake-up distance.
                }
                 \label{fig:SynchronizeDistance}
        \end{minipage}
        \begin{minipage}{0.38\linewidth}  
                \centering
                \vspace{-0.1cm}
                \setlength{\abovecaptionskip}{0.cm}
                \setlength{\belowcaptionskip}{0.cm}
                \includegraphics[width=0.8\linewidth]{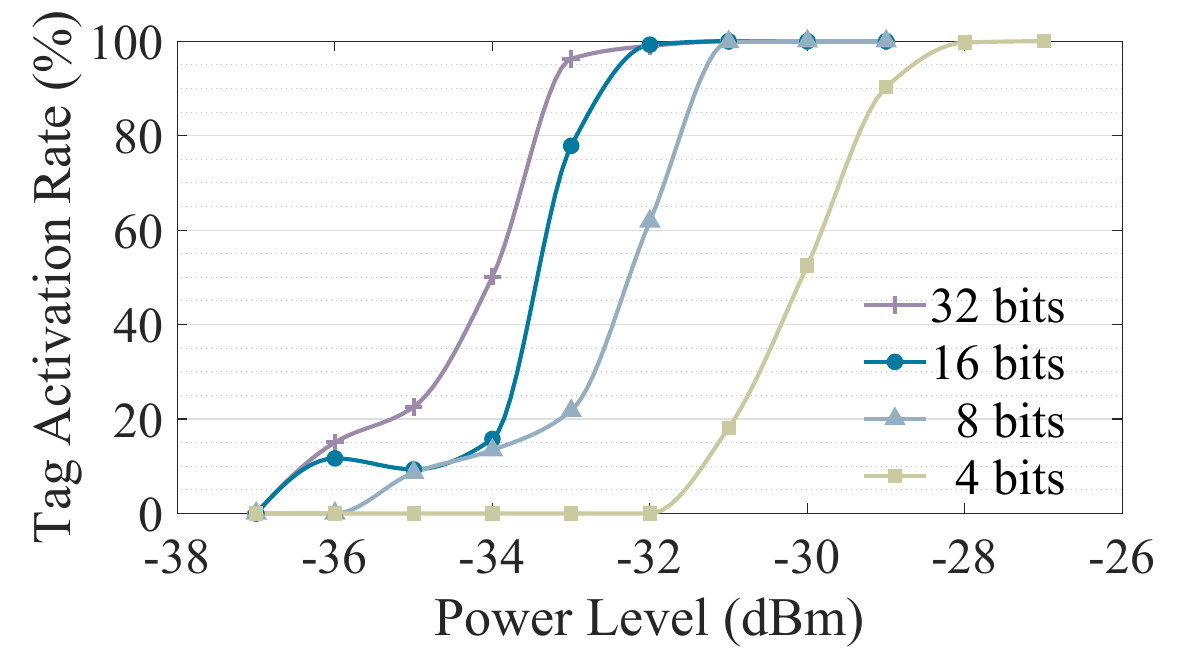}
                \caption{Tag activation rate.}
                \label{fig:ActivateMultipleTag}
        \end{minipage}
        \begin{minipage}{0.34\linewidth}  
                \centering
                \setlength{\abovecaptionskip}{0.cm}
                \setlength{\belowcaptionskip}{0.cm}
                \includegraphics[width=\linewidth]{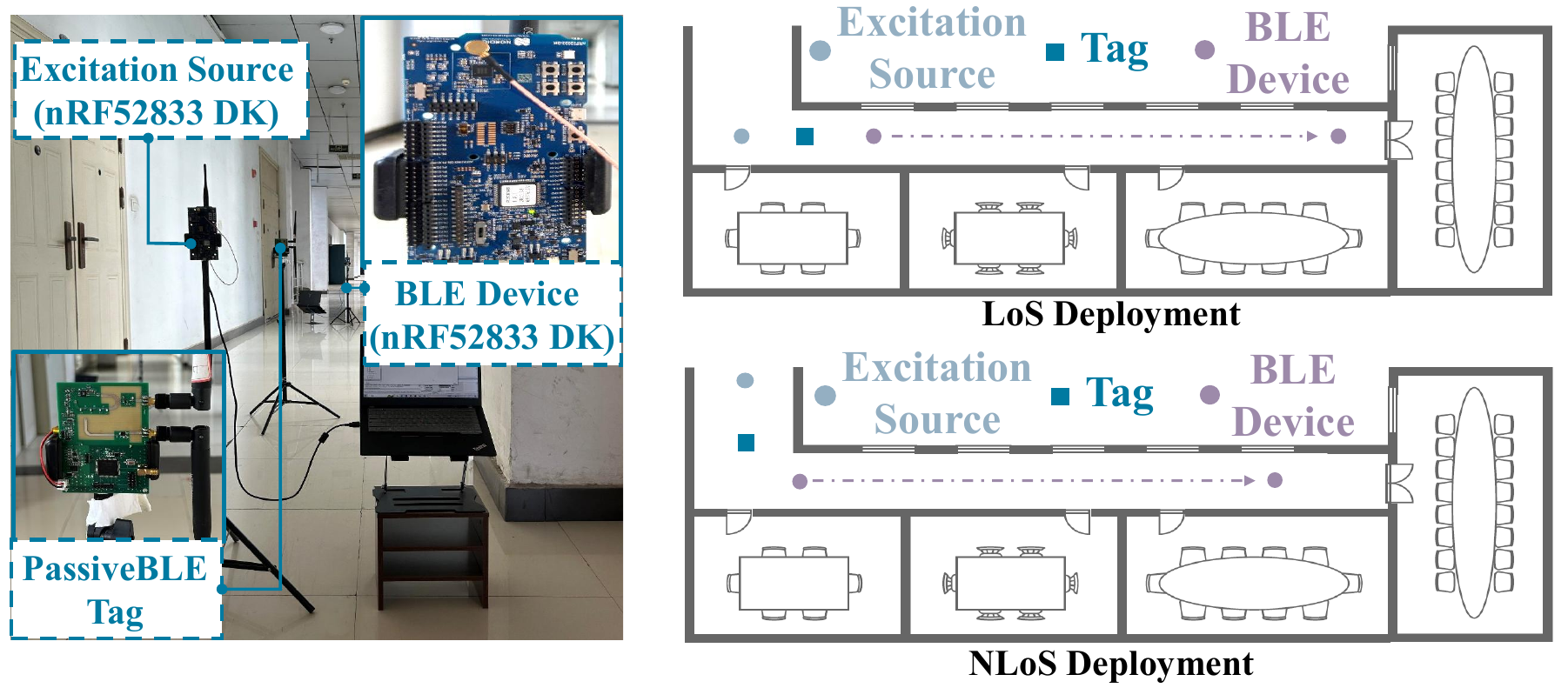}
                \caption{Experimental setup.}
                \label{fig:ExperimentalSet}
        \end{minipage}
 \vspace{-0.3cm}
\end{figure*}

\begin{figure*}[t]
    \centering
    \setlength{\abovecaptionskip}{0.cm}
    \setlength{\belowcaptionskip}{0.cm}
    \subfigure[Goodput.]{
			\includegraphics[width=0.31\linewidth]{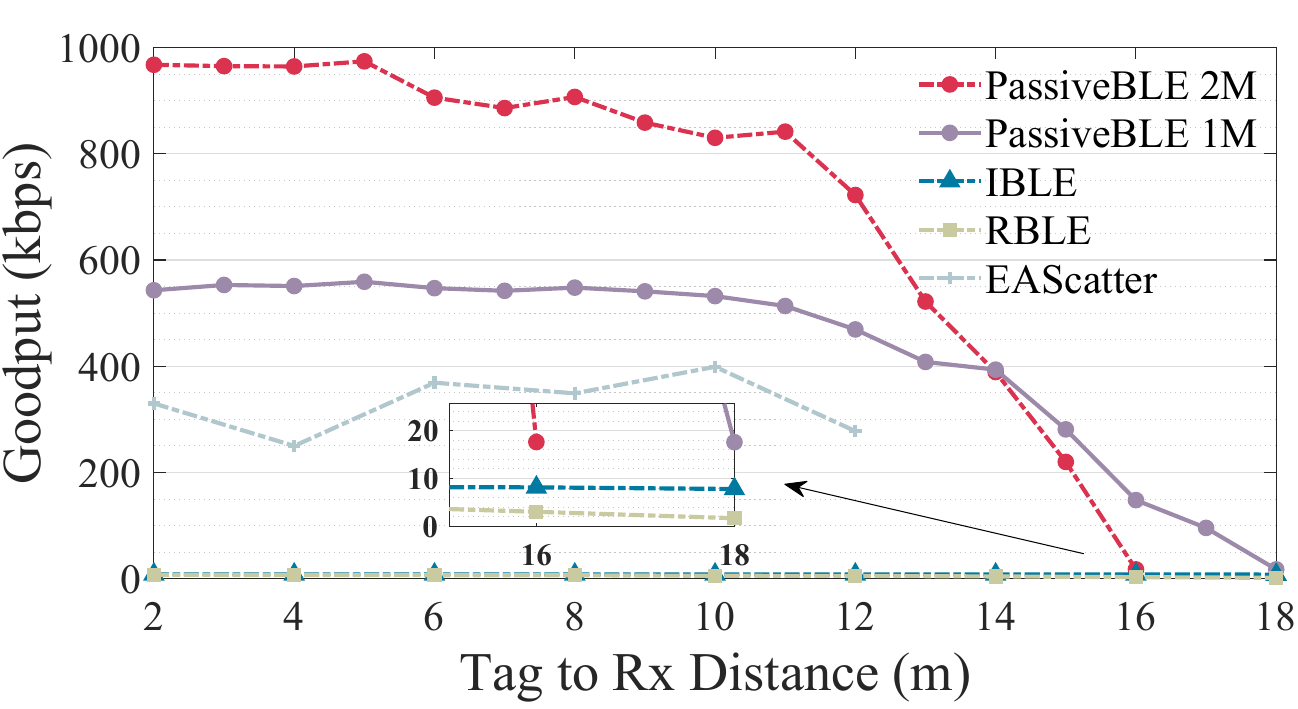}
          }
     \subfigure[BER.]{
			\includegraphics[width=0.31\linewidth]{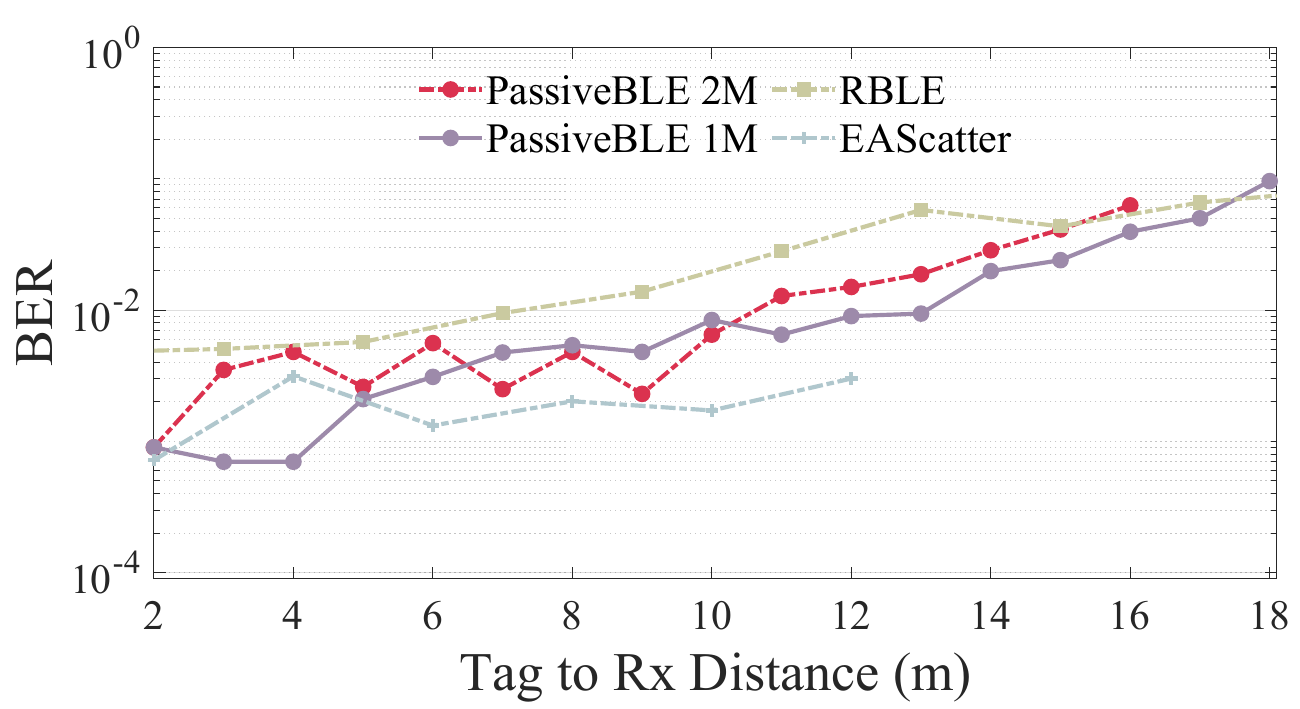}
          }
     \subfigure[RSS.]{
			\includegraphics[width=0.31\linewidth]{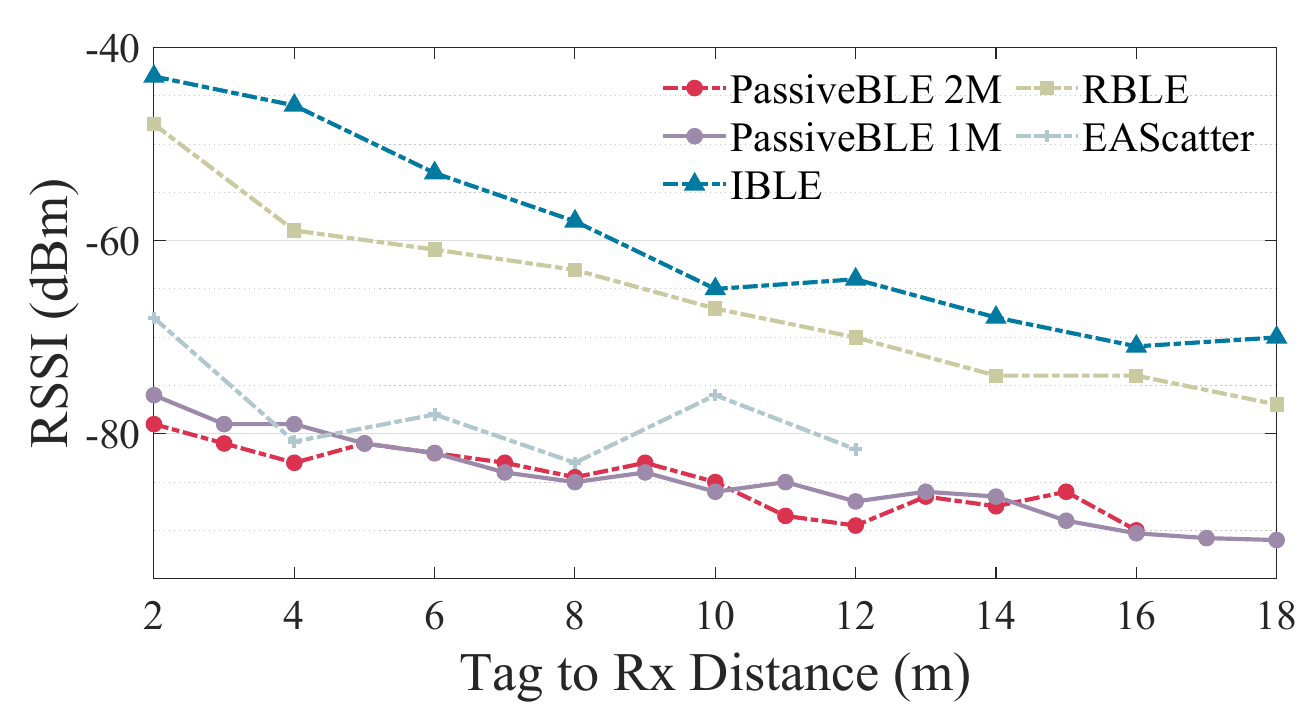}
          }
    \caption{End-to-end communication performance with increasing distance in the LoS condition.}
    \label{fig:LoSPerformance}
    \vspace{-0.3cm}
\end{figure*}

\section{Evaluation}\label{evaluation}
In this section, we first evaluate the performance of the synchronization circuit, including the accuracy and the sensitivity. Then, we evaluate the end-to-end communication performance of \sysname system in both NLoS and LoS scenarios. Finally, we conduct comprehensive experiments to evaluate the \sysname performance on establishing and maintaining BLE connections.

\subsection{Synchronization Performance}
Synchronization performance is crucial to support the symbol-to-symbol XOR operation and quick response to maintain the BLE connections. Here we verify the tag's synchronization performance with micro-benchmark experiments to show the wake-up, activation rate and synchronization performance.
We use two metrics to evaluate the tag's sensitivity: wake-up rate and activation rate. The wake-up rate represents the tag's ability to achieve symbol-level synchronization with the excitation source’s data packet. The activation rate refers to the tag's capability to detect its own ID address from the pre-modulated BLE packets generated by the excitation source.
Additionally, we define synchronization jitter as the accuracy of symbol alignment between the backscatter tag and the excitation source.

\textbf{Wake-up sensitivity.} To evaluate the performance of the sensitivity to wake up a \sysname tag, we configure a BLE transceiver as the excitation source to send BLE data packets with power levels from -40~dBm to -25~dBm. The excitation source and the tag are connected with a cable to reduce the varying wireless channel's influence on the results. If the tag has been woken up by the data packet, it will send a trigger signal. 
At each power level, we send 10,000 data packets and calculate the wake-up rate by dividing the count of trigger signals by the total packets sent. The results are plotted in \fig~\ref{fig:Synchronization_accuracy}(a). The sensitivity is about -30~dBm where the successful rate is about 91\%. When the power level falls below -31~dBm, the successful rate rapid drops to 24\%. 

\textbf{Tag activation rate.} As stated in \S~3.2, we can use redundant bits in the \textit{Access Address} part to improve the sensitivity of tag activation. We conduct a micro-benchmark experiment with the same setup to evaluate the tag activation rate with different bits.  
As plotted in \fig~\ref{fig:ActivateMultipleTag}, the more redundant bits lead to a better tag activation rate with low power level BLE data packets. Since the wake-up and synchronization sensitivity with the BLE preamble is about -30~dBm, we choose $n=8$ to balance the wake-up sensitivity and the number of tags, where $2^{32/8}=16$ tags are sufficient for some BLE backscatter applications in wearable or smart homes. 

\textbf{Synchronization jitter.} 
To quantify synchronization accuracy, we use synchronization jitter, as is standard in existing backscatter synchronization systems~\cite{dunna2021syncscatter}. Jitter represents the time difference or errors between the symbol rising edges of the excitation source and the tag. We use a setup similar to the previous system~\cite{dunna2021syncscatter} to measure the time jitter but configure the excitation source to pull up the voltage of an onboard input-output (IO) interface to mark the beginning of the packet. The tag will also pull up the voltage of an IO interface after it successfully detects the preamble. We use an oscilloscope to monitor the delay of the voltage pull-up edge between these two IO ports to acquire the synchronization latency and record the jitter of the edge of the tag's IO interface. The average delay from the transceiver sending the data packets to the \sysname tag detecting the signal is about 10.4~$\mu$s, where the preamble lasts for 8~$\mu$s and the onboard processing costs 2.4~$\mu$s.  
The jitter at different power levels is plotted in \fig~\ref{fig:Synchronization_accuracy}(b). The jitter is below 23.7~ns with a power level above -30~dBm and 93.1~ns with a power level above -32~dBm, which achieves the desired optimal spec for ns-level synchronization accuracy. 
\begin{figure*}[t]
    \centering
    \setlength{\abovecaptionskip}{0.cm}
    \setlength{\belowcaptionskip}{0.cm}
    \subfigure[Goodput.]{
			\includegraphics[width=0.32\linewidth]{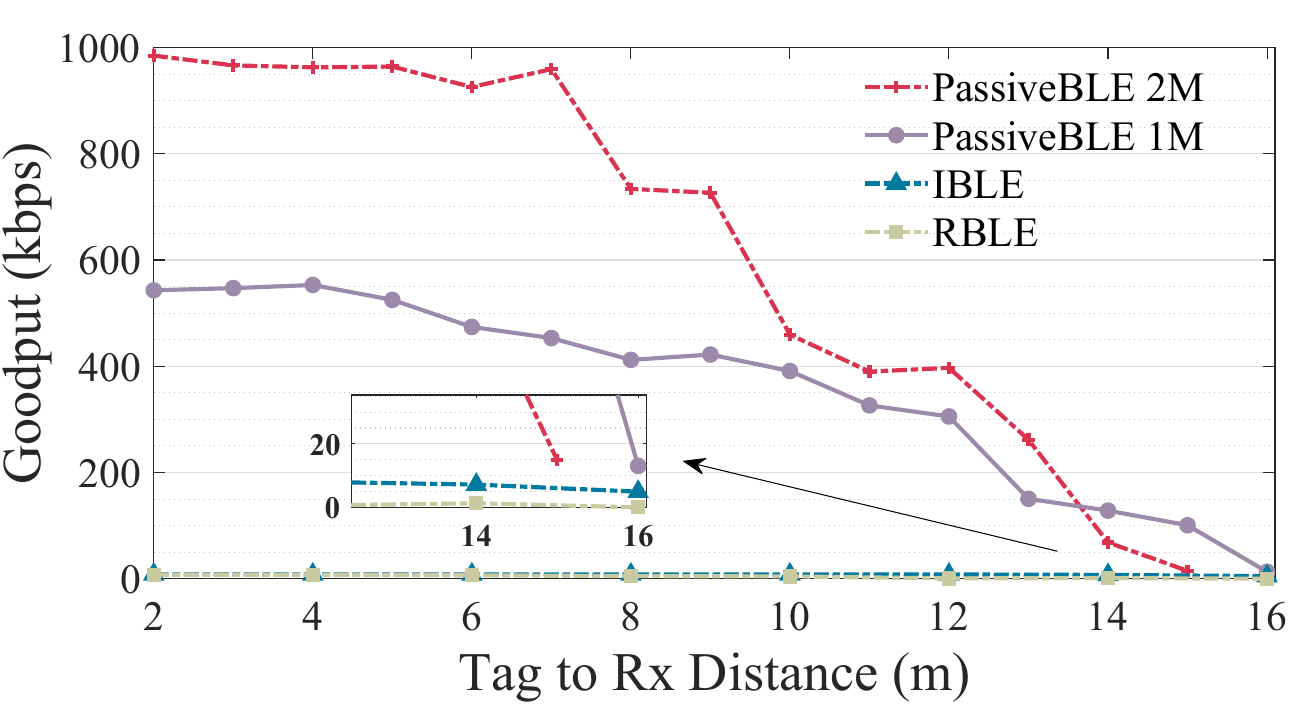}
          }
     \subfigure[BER.]{
			\includegraphics[width=0.32\linewidth]{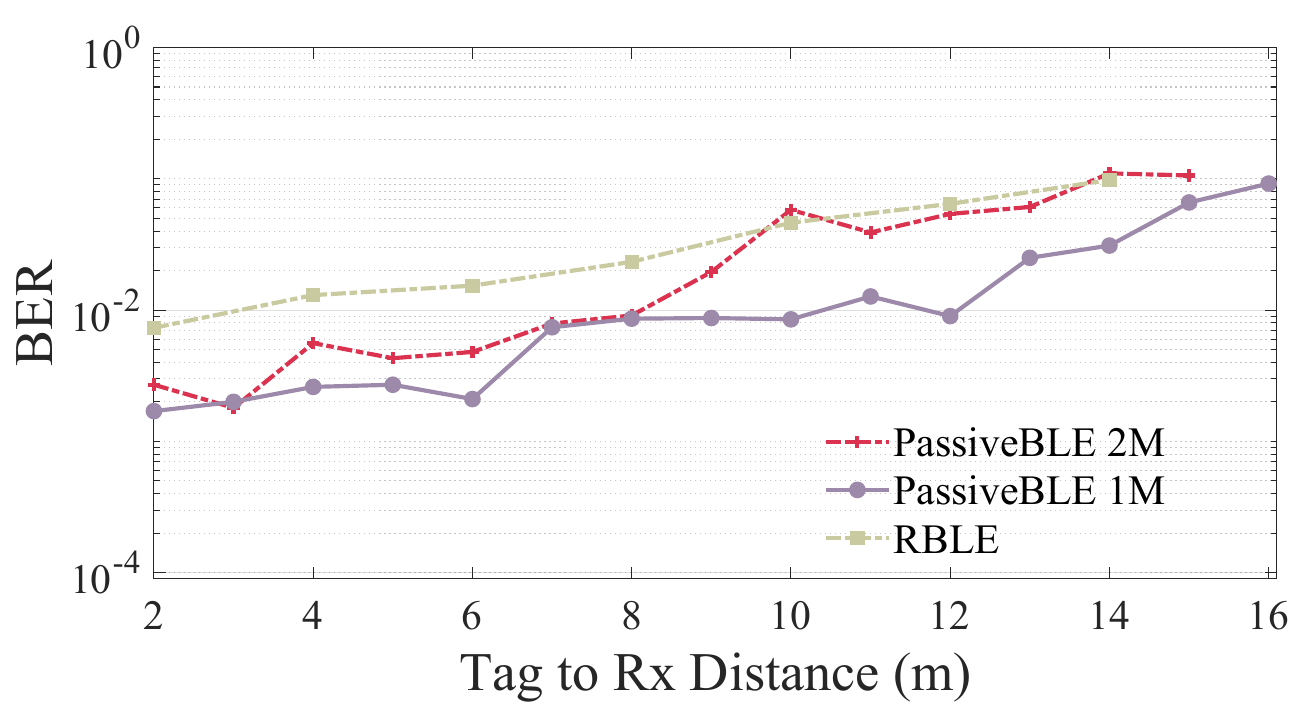}
          }
     \subfigure[RSS.]{
			\includegraphics[width=0.32\linewidth]{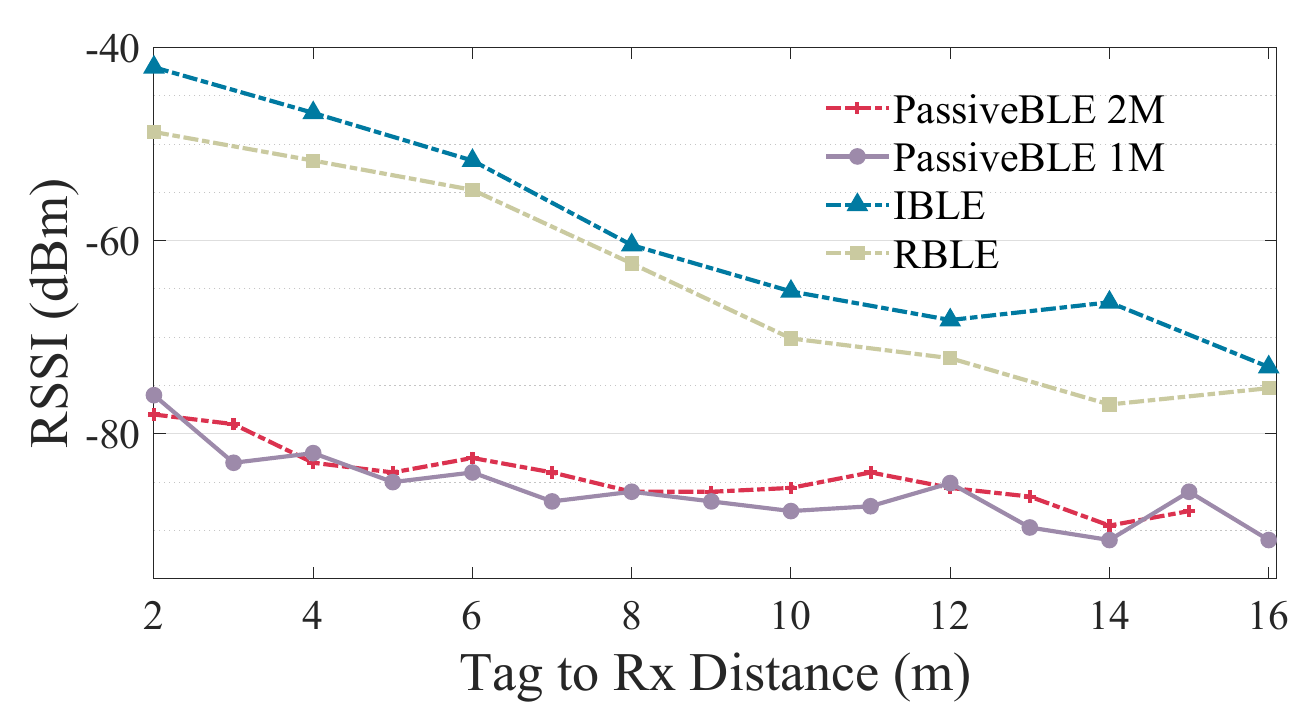}
          }
    \caption{End-to-end communication performance with increasing distance in the NLoS condition.}
    \label{fig:NLoSPerformance}
      \vspace{-0.2cm}
\end{figure*}
\begin{figure*}[t]
        \begin{minipage}{0.33\linewidth}  
                \centering
                \vspace{-0.1cm}
                \setlength{\abovecaptionskip}{0.cm}
                \setlength{\belowcaptionskip}{0.cm}
                \includegraphics[width=\linewidth]{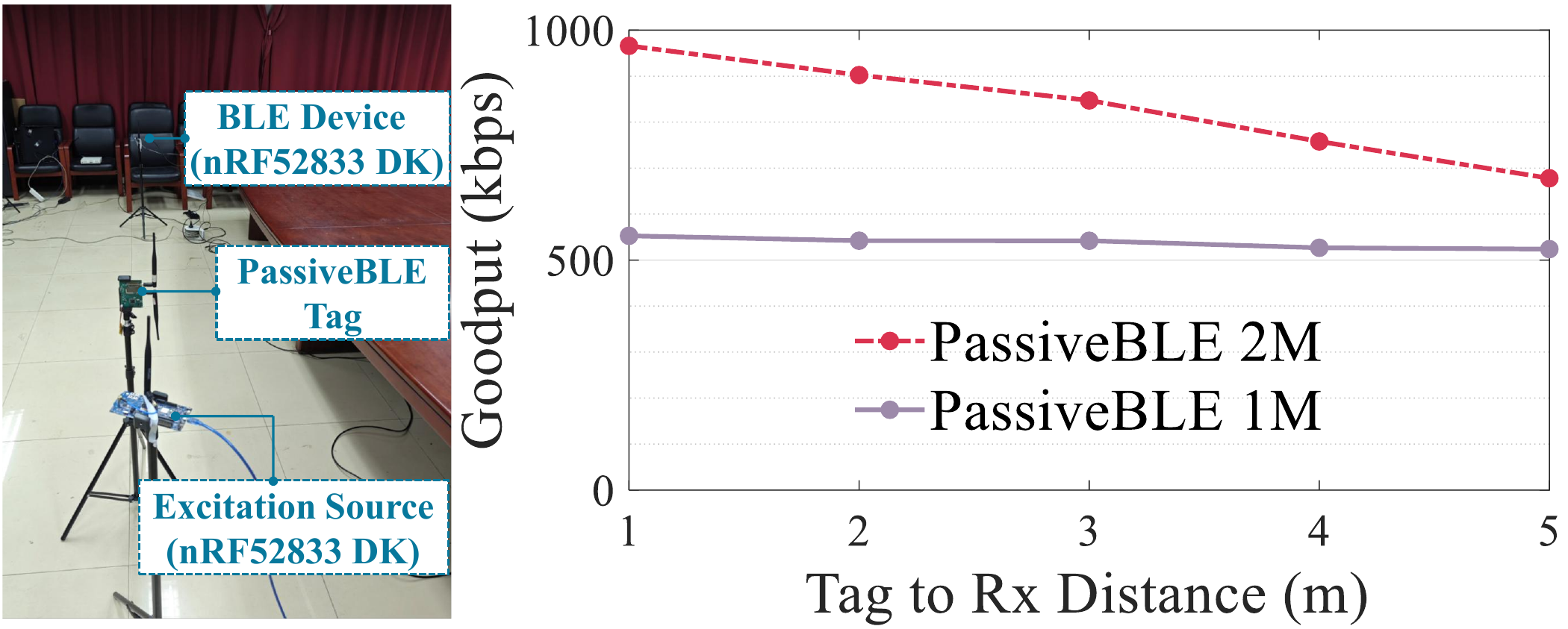}
                \caption{Communication \\performance in a meeting room.}
                 \label{fig:room401}
        \end{minipage}
        \begin{minipage}{0.33\linewidth}  
                \centering
                \setlength{\abovecaptionskip}{0.cm}
                \setlength{\belowcaptionskip}{0.cm}
                \includegraphics[width=\linewidth]{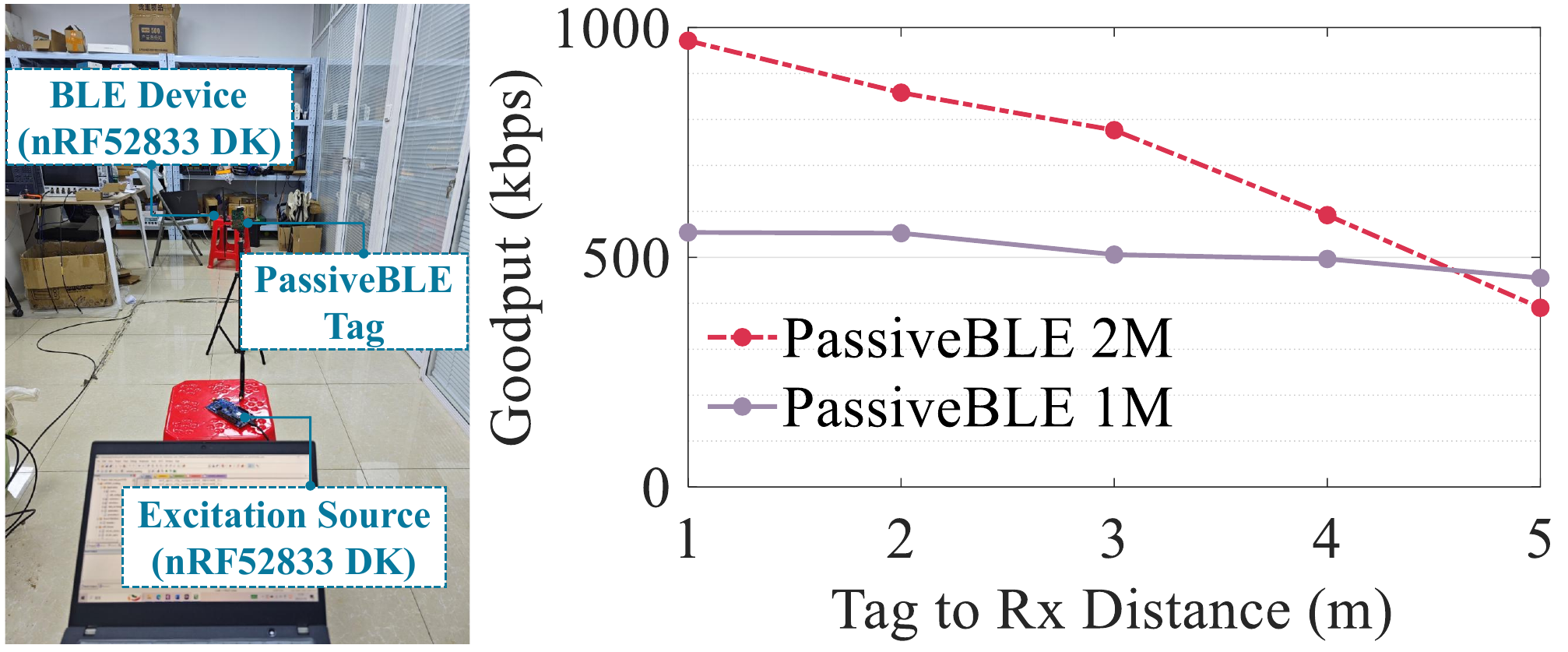}
                \caption{Communication \\ Performance in a cluttered lab.}
                \label{fig:room403}
        \end{minipage}
        \begin{minipage}{0.33\linewidth}  
                \centering
                \vspace{-0.1cm}
                \setlength{\abovecaptionskip}{0.cm}
                \setlength{\belowcaptionskip}{0.cm}
                \includegraphics[width=\linewidth]{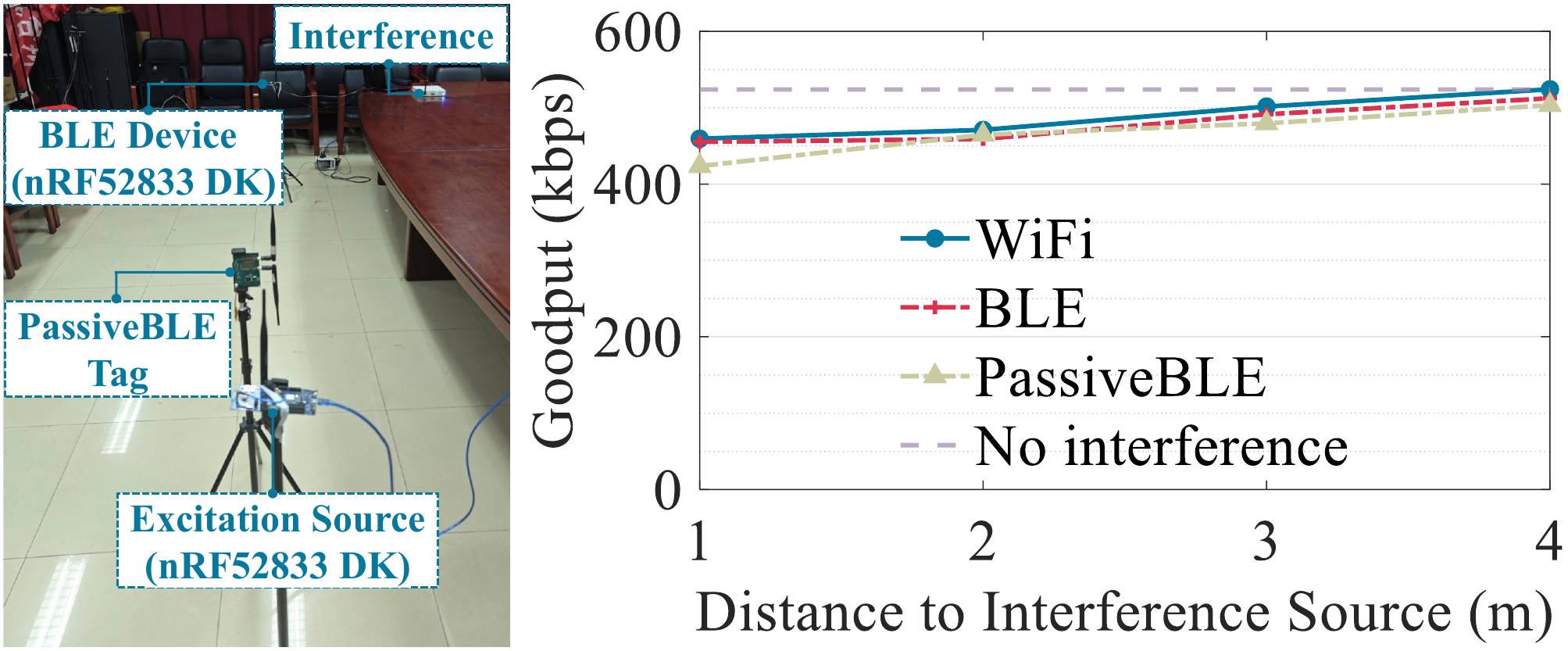}
                \caption{Performance under \\  different interference sources.}
                \label{fig:DiffInterferSources}
        \end{minipage}
\vspace{-0.3cm}
\end{figure*}

\textbf{Synchronization distance.}
We fix a BLE excitation source at the end of a corridor with EIRP of the excitation source at 20~dBm, then move a \sysname tag along the corridor with increasing distances away from the excitation source from 0 m to 5 m with 0.25~m increments. We configure the excitation source to transmit a dedicated BLE packet with a standard preamble and the tag's address is loaded in the \textit{ACCESS ADDRESS} part. At each distance, the excitation source transmits ten thousands BLE packets to trigger the \sysname tags, and we record the times that successfully wake up and trigger the tag. 
As plotted in \fig~\ref{fig:SynchronizeDistance}, results show that the success rate keeps above 80.0\% with the distance increase to 2.5~meters and above 95.1\% when the distance to the excitation source is within 1.5~meters.

\vspace{-0.3cm}
\subsection{Communication Performance} 	

\textbf{Experimental setup.} As shown in \fig\ref{fig:ExperimentalSet}, we evaluate the performance of \sysname in both line-of-sight (LoS) and non-line-of-sight (NLoS) scenarios. We use the TX-to-Tag distance to refer the distance from an excitation source to a \sysname tag, and the Tag-to-RX distance to refer the distance from a \sysname tag to the receiver. 
The default TX-to-Tag distance is set to 0.5~m, the same as the previous BLE backscatter system~\cite{zhang2021commodity} for a fair comparison. In the LOS scenarios, we place the excitation source, tags, and receivers in the corridor, while in the NLOS scenarios, the excitation source and tags are placed at the corner of the corridor. 
We move the receiver along a straight line to increase the distance of Tag-to-RX distance and record the received packets for evaluation.
In each location, we employ the BLE transceiver to receive packets for 5 minutes and calculate the average performance. Then compare our results with the SOTA solutions, including IBLE~\cite{zhang2021commodity}, EAScatter~\cite{EAScatter} and RBLE~\cite{zhang2020reliable}, with the published performance in their papers~\cite{zhang2021commodity, EAScatter}. In particular, IBLE and RBLE are compared in both LoS and NLoS scenarios and EAScatter~\cite{EAScatter} is compared in LoS. Some BLE backscatter systems~\cite{jiang2023dances,jiang2023bidirectional,huang2024bitalign} are excluded because of their different setup or lack of similar experiments, their measured goodput is listed in the Table~\ref{Table:ImplementPower}.

We test the performance of \sysname tags in both of the \textit{LE 1M PHY} and \textit{LE 2M PHY} modes in these two scenarios. The results are plotted in \fig~\ref{fig:LoSPerformance} and \fig~\ref{fig:NLoSPerformance}. Overall, \sysname achieves remarkable performance improvement in goodput and bit error rate (BER) with a lower receive signal strength (RSS). 
\sysname achieves full compatibility with unmodified BLE transceivers and reaches a goodput of up to 532 ~kbps in \textit{LE 1M PHY} mode, which is 63.3$\times$ higher than that of the existing commodity-level BLE backscatter~\cite{zhang2021commodity}, in the LoS scenario.
Although the goodput of the \sysname tag drops quickly when the distance exceeds ten meters, \sysname still realizes 3$\times$ higher goodput than the commodity compatible IBLE system. Our experiment confirms that \sysname outperforms the SOTA commodity-compatible backscatter system~\cite{zhang2021commodity} in terms of goodput.
We now elaborate on the performance in LoS and NLoS scenarios, respectively. 

\begin{figure*}[t]
    \hfill
    \centering
    \setlength{\abovecaptionskip}{0.cm}
    \setlength{\belowcaptionskip}{0.cm}
        \begin{minipage}{0.485\linewidth}  
          \subfigure[Goodput.]{
                                \includegraphics[width=0.48\linewidth]{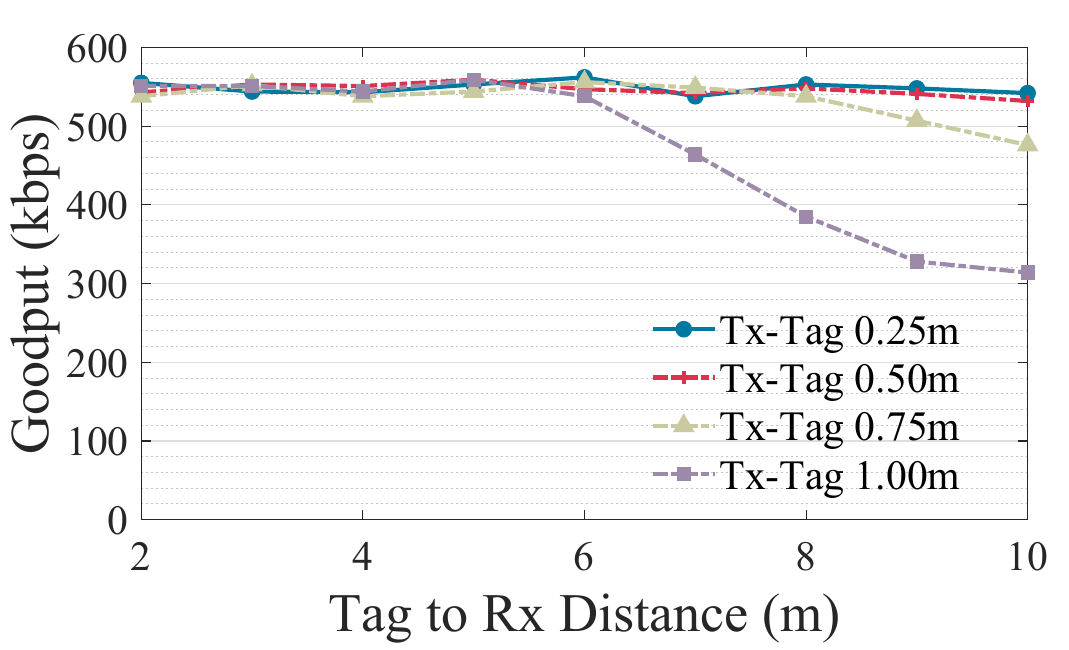}}
          \subfigure[BER.]{
                                \includegraphics[width=0.48\linewidth]{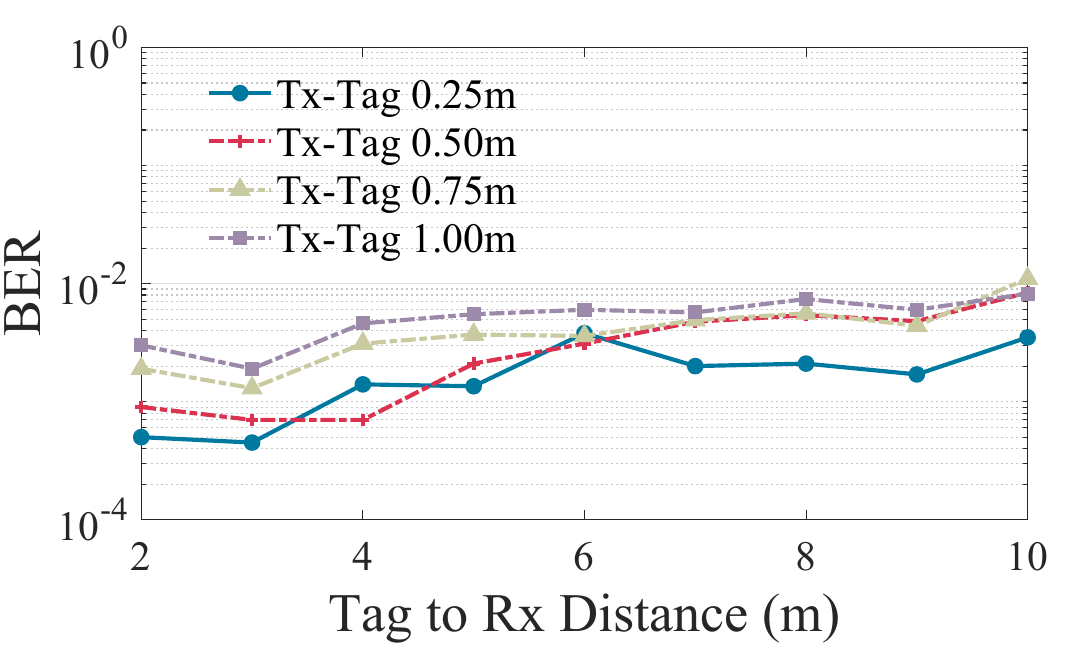}}
        \caption{Performance versus different distances in LE 1M Mode.}
        \label{fig:Downlink_1M_Performance}                       
        \end{minipage}
        \hspace{0.1cm}
        \begin{minipage}{0.49\linewidth}  
                        \subfigure[Goodput.]{
                                \includegraphics[width=0.47\linewidth]{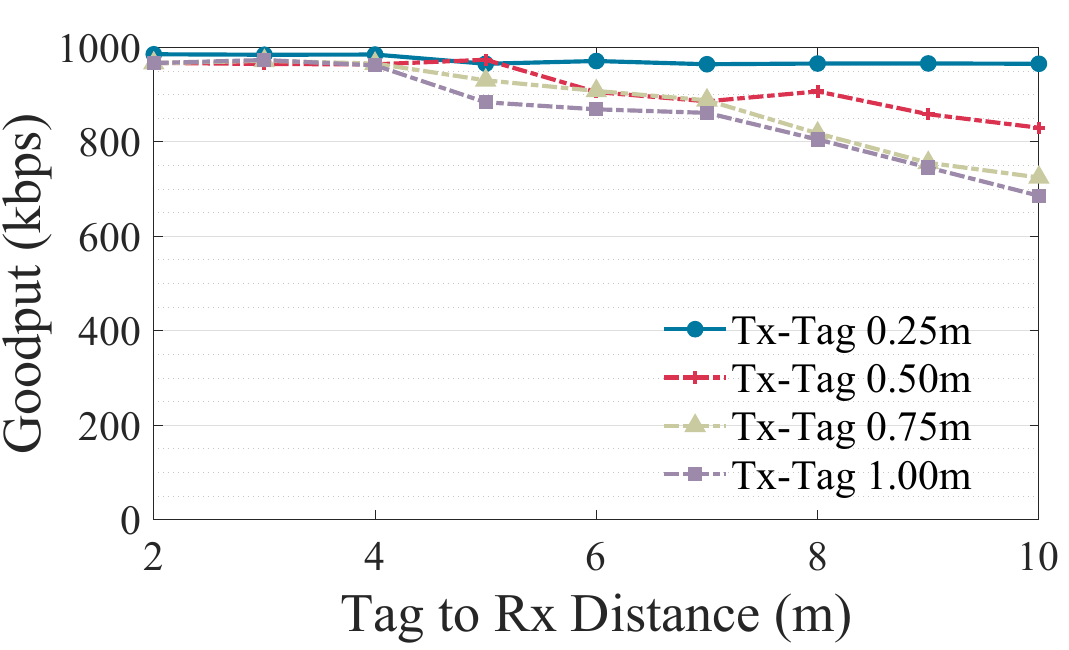}}
                        \subfigure[BER.]{
                                \includegraphics[width=0.47\linewidth]{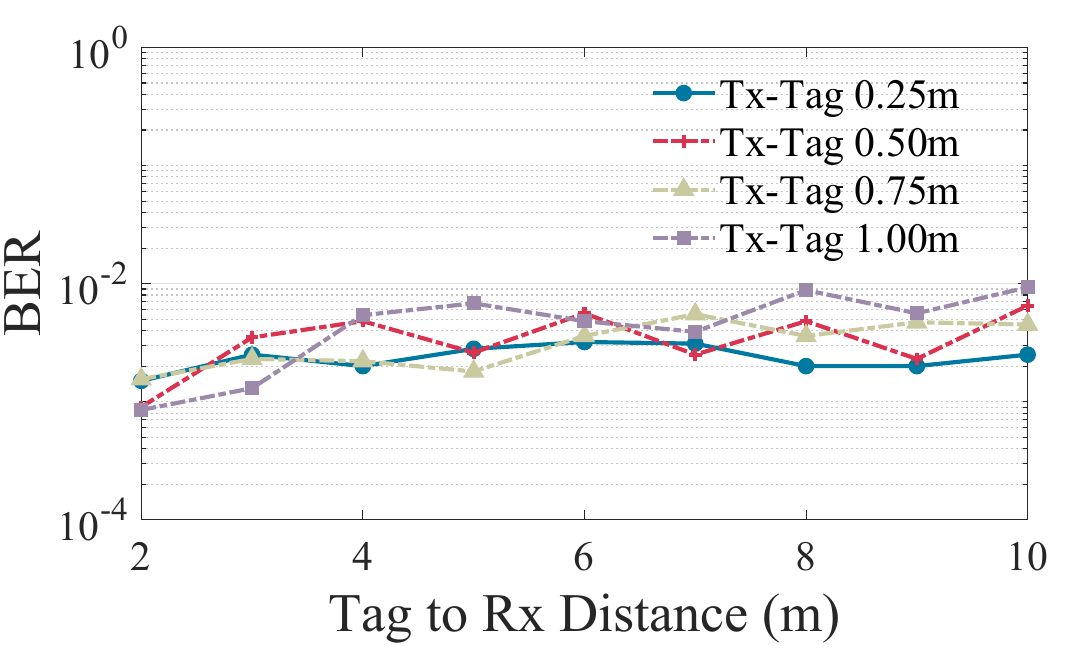}}
                \caption{Performance versus different distances in LE 2M Mode.}
                \label{fig:Downlink_2M_Performance}
        \end{minipage}
 \vspace{-0.3cm}
\end{figure*}

\begin{figure*}[t]
    \hfill
        \begin{minipage}{0.33\linewidth}  
                \centering
                \vspace{-0.1cm}
                \setlength{\abovecaptionskip}{0.cm}
                \setlength{\belowcaptionskip}{0.cm}
                \includegraphics[width=\linewidth]{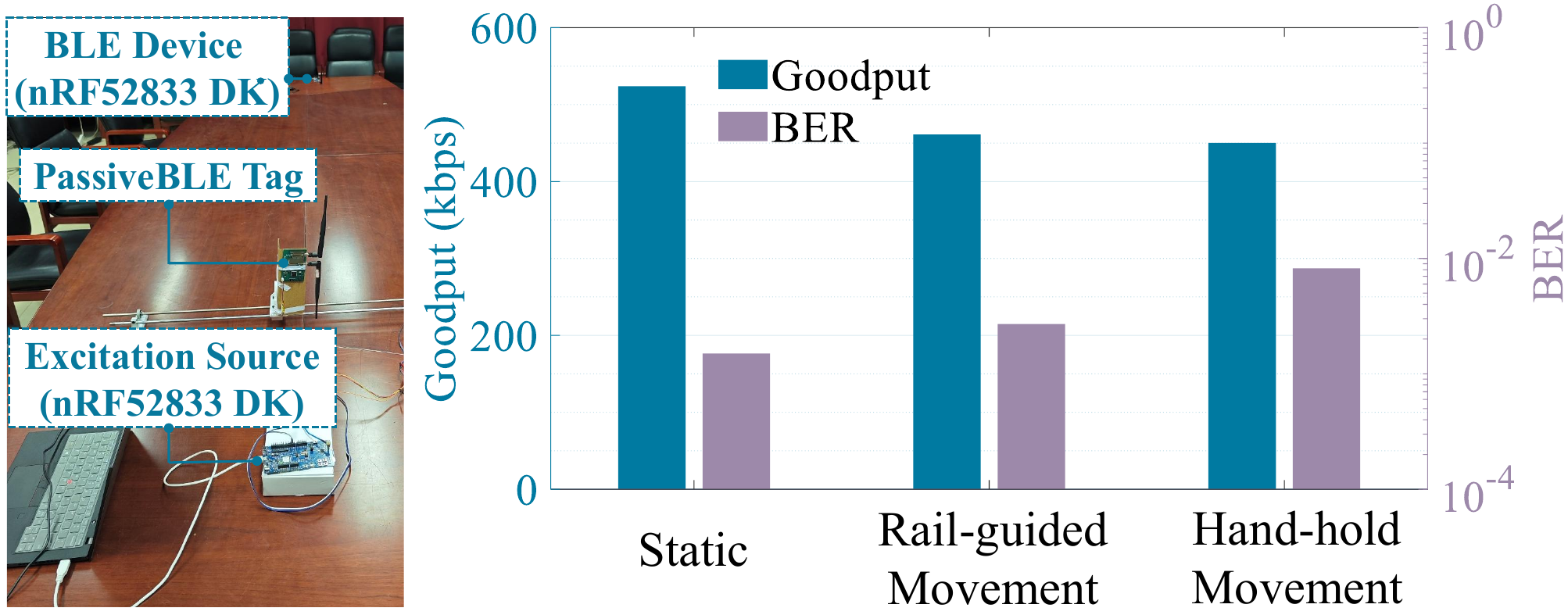}
                \caption{Performance under \\ different movement.
                }
                 \label{fig:differMove}
        \end{minipage}
        \begin{minipage}{0.32\linewidth}  
                \centering
                \vspace{-0.1cm}
                \setlength{\abovecaptionskip}{0.cm}
                \setlength{\belowcaptionskip}{0.cm}
                \includegraphics[width=\linewidth]{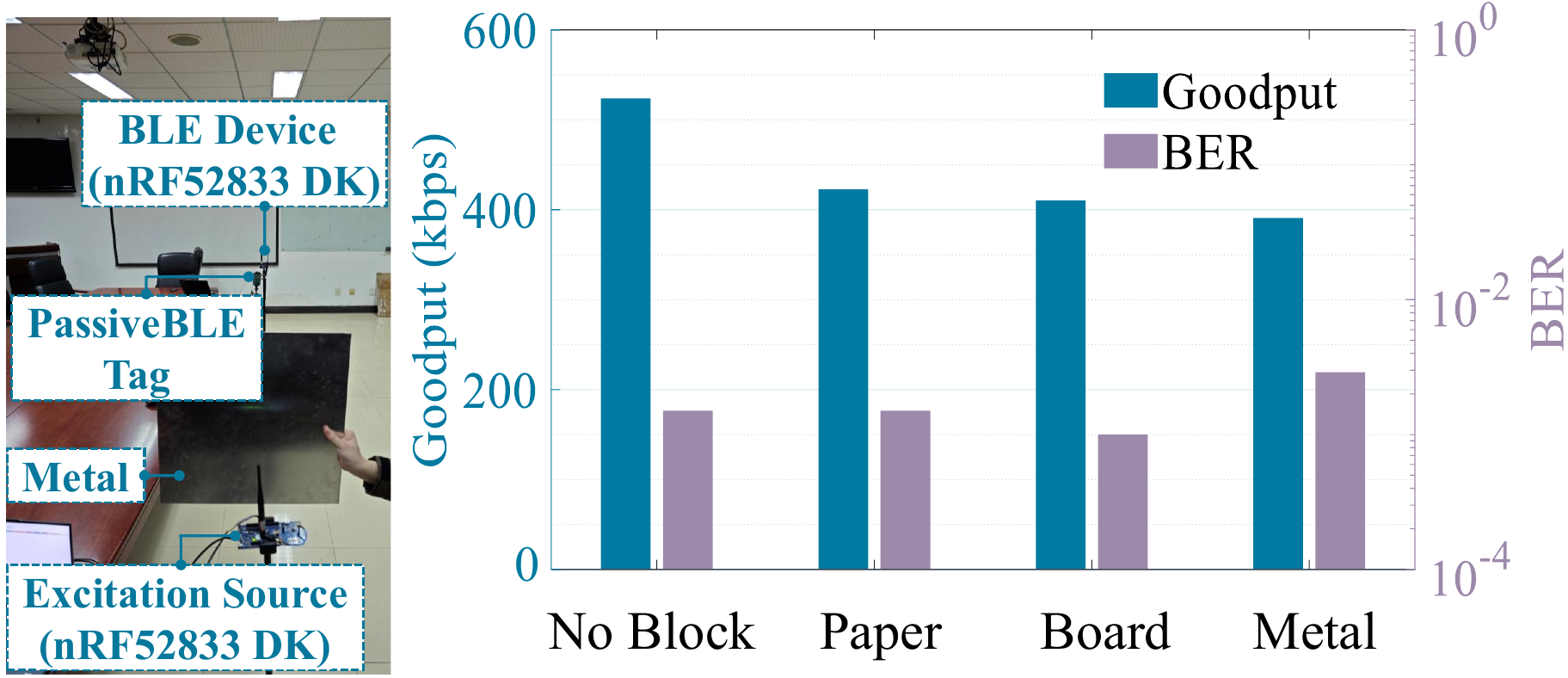}
                \caption{Performance with \\ different blocks.}
                \label{fig:differBlock}
        \end{minipage}
        \begin{minipage}{0.32\linewidth}  
                \centering
                \setlength{\abovecaptionskip}{0.cm}
                \setlength{\belowcaptionskip}{0.cm}
                \includegraphics[width=\linewidth]{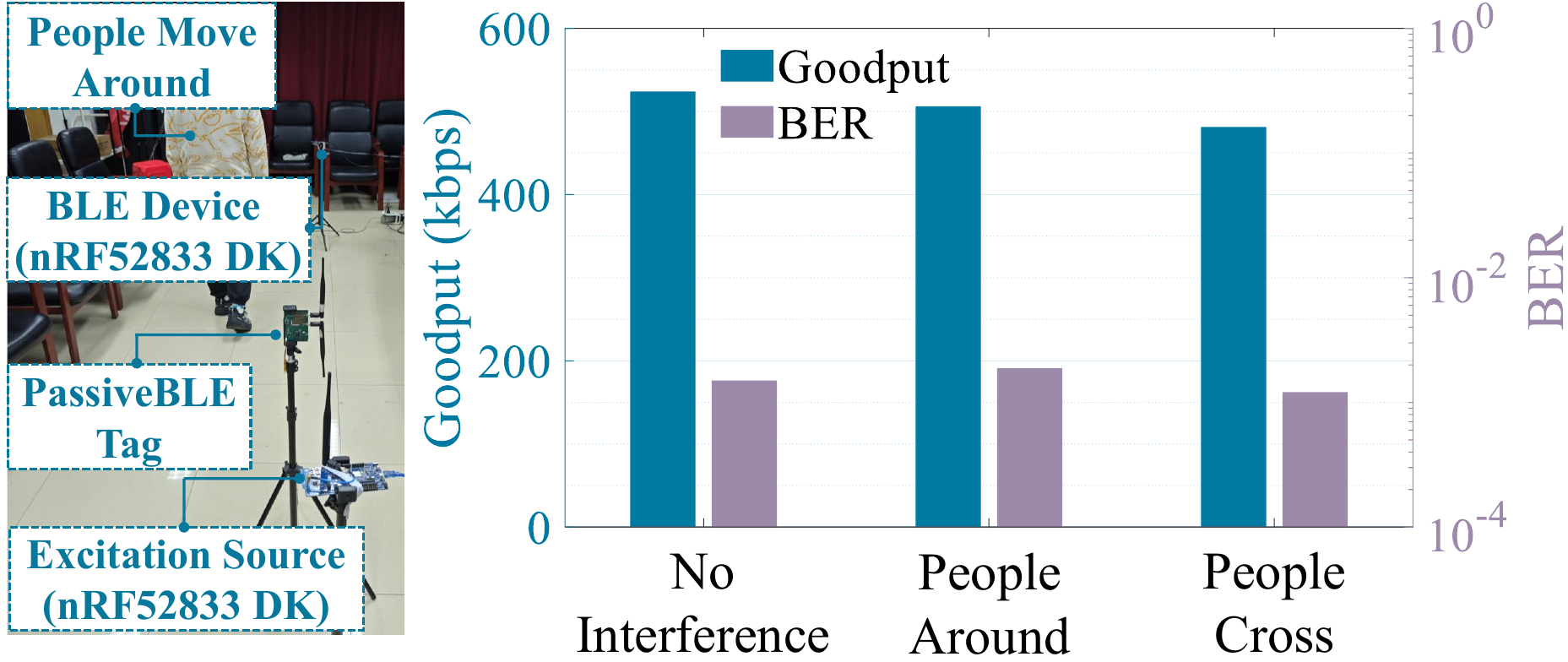}
                \caption{Performance under \\ interference with nearby people.}
                \label{fig:peopleInfer}
        \end{minipage}
 \vspace{-0.3cm}
\end{figure*}

\textbf{LoS scenario.} As plotted in \fig~\ref{fig:LoSPerformance}, the performance of \sysname is superior from both of the goodput and RSS aspects in most distances. 
In LoS scenarios, \sysname achieves a high goodput up to 532~kbps in \textit{LE 1M PHY} mode and 974~kbps in \textit{LE 2M PHY} mode, which achieves orders of goodput improvement than existing BLE backscatter systems~\cite{zhang2021commodity,zhang2020reliable,zhang2017freerider,jiang2023dances,jiang2023bidirectional} and comparable to active BLE chips in most BLE applications~\cite{AmbientIoT_Market_research, AmbientIoT_BLE_vehicle}. 
Furthermore, as plotted in \fig\ref{fig:LoSPerformance}~(b), \sysname presents lower RSS performance to achieve the higher goodput, which indicates the \sysname is more power efficient. 
We also conduct experiments in two indoor scenarios, a meeting room and a cluttered lab, both of which tend to suffer from severe multipath interference, to evaluate the communication performance in more complex environments. Results are plotted in Figures~\ref{fig:room401} and~\ref{fig:room403}. The goodput of PassiveBLE tags indeed degrades in these environments, with the degradation in LE 2M mode being more significant. Nonetheless, the system can still maintain about 500 kbps (in the meeting room) or 480 kbps (in the clutter lab) within 5 meters, thereby meeting the application requirements in these environments.

\textbf{NLoS scenario.} As plotted in \fig~\ref{fig:NLoSPerformance}, \sysname also achieves a superior goodput when the Tag-to-RX distance is below ten meters in NLoS scenarios. The measured average goodput of \sysname tag with a Tag-to-RX distance of ten meters is 459.5~kbps in \textit{LE 2M PHY} mode and 391.0~kbps in \textit{LE 1M PHY}, which dropped as expected due to the occlusion.  However, the performance of \sysname is still sufficient for many IoT applications, e.g., sensor data collection~\cite{huang2018vehicle} and audio streaming~\cite{LEAudio}.

\begin{figure}[t]
    \centering
            \setlength{\abovecaptionskip}{0.1cm}
            \setlength{\belowcaptionskip}{0.1cm}
                \begin{minipage}[t]{0.48\linewidth}
                        \raggedleft
                        \subfigure[Establishment performance.]{
                                \includegraphics[width=\linewidth]{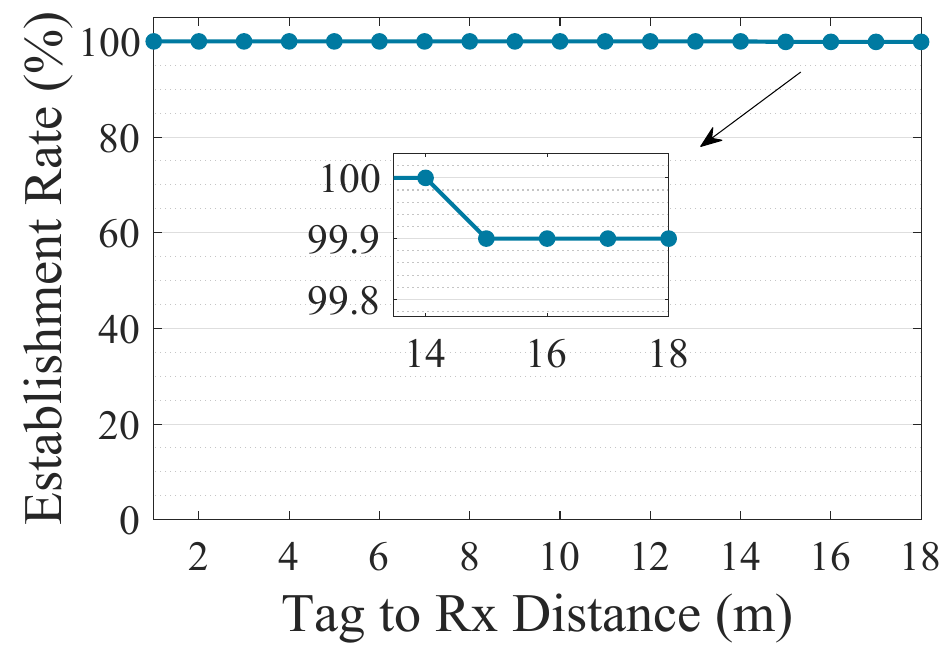}}
                \end{minipage}
                 \hfill
                \begin{minipage}[t]{0.48\linewidth}
                        \raggedleft
                        \subfigure[Maintenance performance.]{
                                \includegraphics[width=\linewidth]{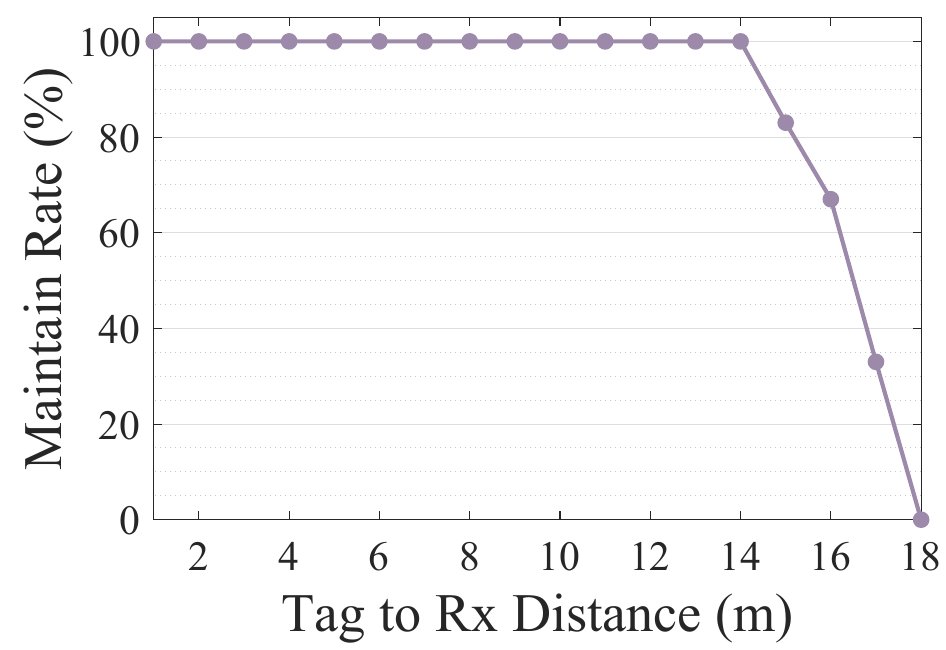}}
                \end{minipage}
                \caption{Connection performance.}
                \label{fig:Establishment_Maintance}
                 \vspace{-0.4cm}
\end{figure}

\textbf{Impact of interference.} We conduct an experiment to evaluate the performance of the PassiveBLE system under various interference sources, including a WiFi device (implemented with a USRP x310) transmitting OFDM symbols in the 2.4 GHz band, a PassiveBLE tag and BLE devices transmitting on random channels. As shown in Figure~\ref{fig:DiffInterferSources}, the goodput drops by about 20\% when the interference source is within 1 meter, yet it remains superior with ample room for performance enhancement through redundant coding methods. The influence is minimal when the distance is beyond 4 meters, indicating that this approach can meet the requirements of smart home or logistics applications.

\textbf{Impact of Tx-to-Tag distance.}
The Tx-to-Tag distance is also a major concern when deploying \sysname tags in practical applications. To evaluate how this distance influences the performance, we move the location of the tag with the distance increasing from 0.25~m to 1.00~m and send ten thousand packets at each location in both BLE physical modes. The results of \sysname in LE 1M and LE 2M modes are plotted in \fig~\ref{fig:Downlink_1M_Performance} and \fig~\ref{fig:Downlink_2M_Performance}. Overall, when the Tx-to-Tag distance is increased to one meter, \sysname can guarantee a goodput over 310~kbps and 680~kbps in LE 1M and LE 2M mode at the uplink distance of ten meters. 

\textbf{Reliability under dynamic environments.} We evaluate performance in dynamic environments, including scenarios with different movements (\fig~\ref{fig:differMove}), blocking conditions (\fig~\ref{fig:differBlock}), and movements of nearby people (\fig~\ref{fig:peopleInfer}), with the Tag-to-RX distance fixed at 5 meters. Specifically, blocking the LoS between the tag and receiver causes the greatest performance degradation, followed by the effect of different movements, while the movement of nearby people has only a slight influence. Overall, even in the worst-case scenario, when the tag is partially blocked by a metal board, it can still provide a superior goodput of up to 390~kbps, offering attractive application potential for short-range, low-power wireless network applications.

\begin{figure*}[t]
     \centering
        \setlength{\abovecaptionskip}{0.cm}
        \setlength{\belowcaptionskip}{0.cm}
        \begin{minipage}{0.49\linewidth} 
        \subfigure[LE 1M Mode.]{
			\includegraphics[width=0.47\linewidth]{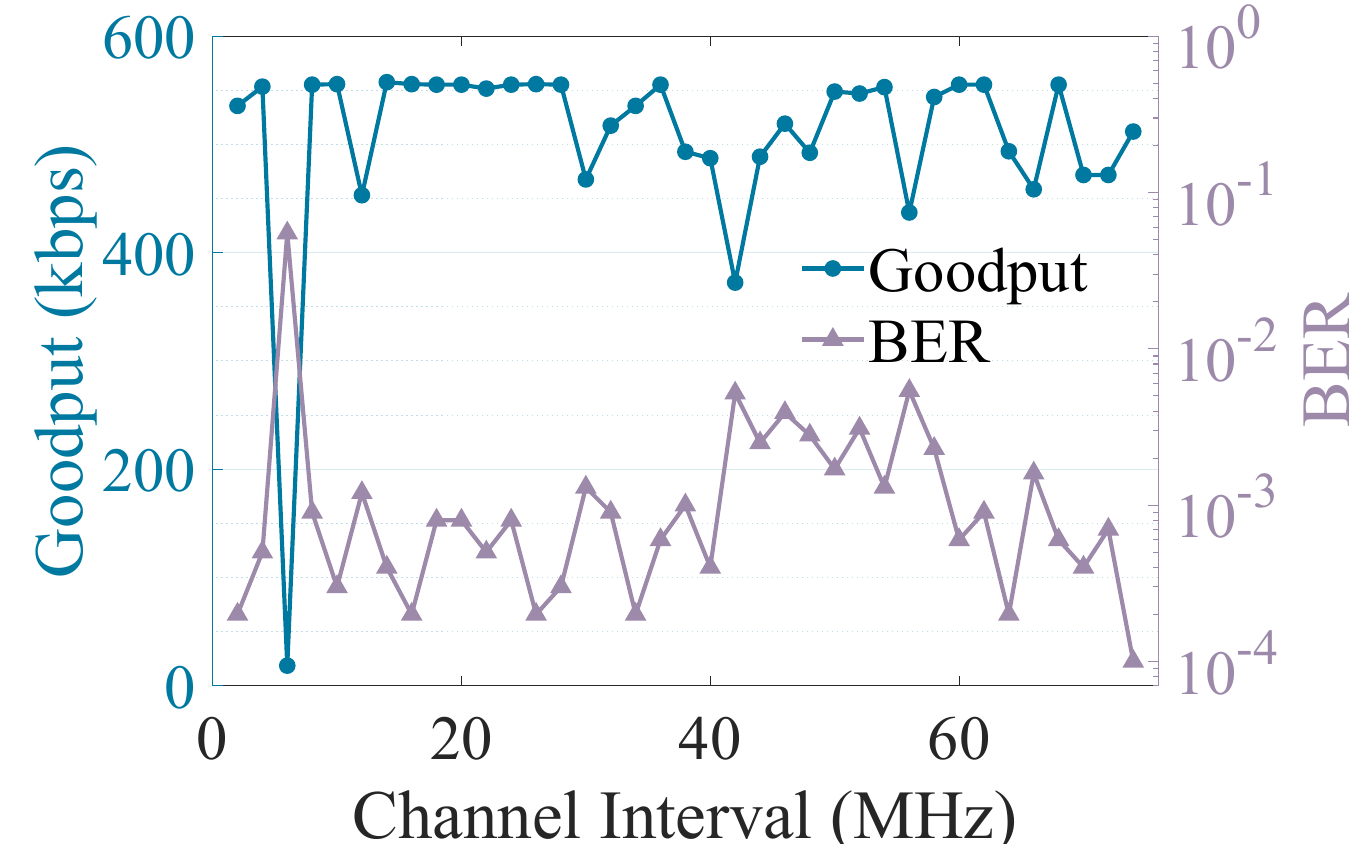}
          }
     \subfigure[LE 2M Mode.]{
			\includegraphics[width=0.47\linewidth]{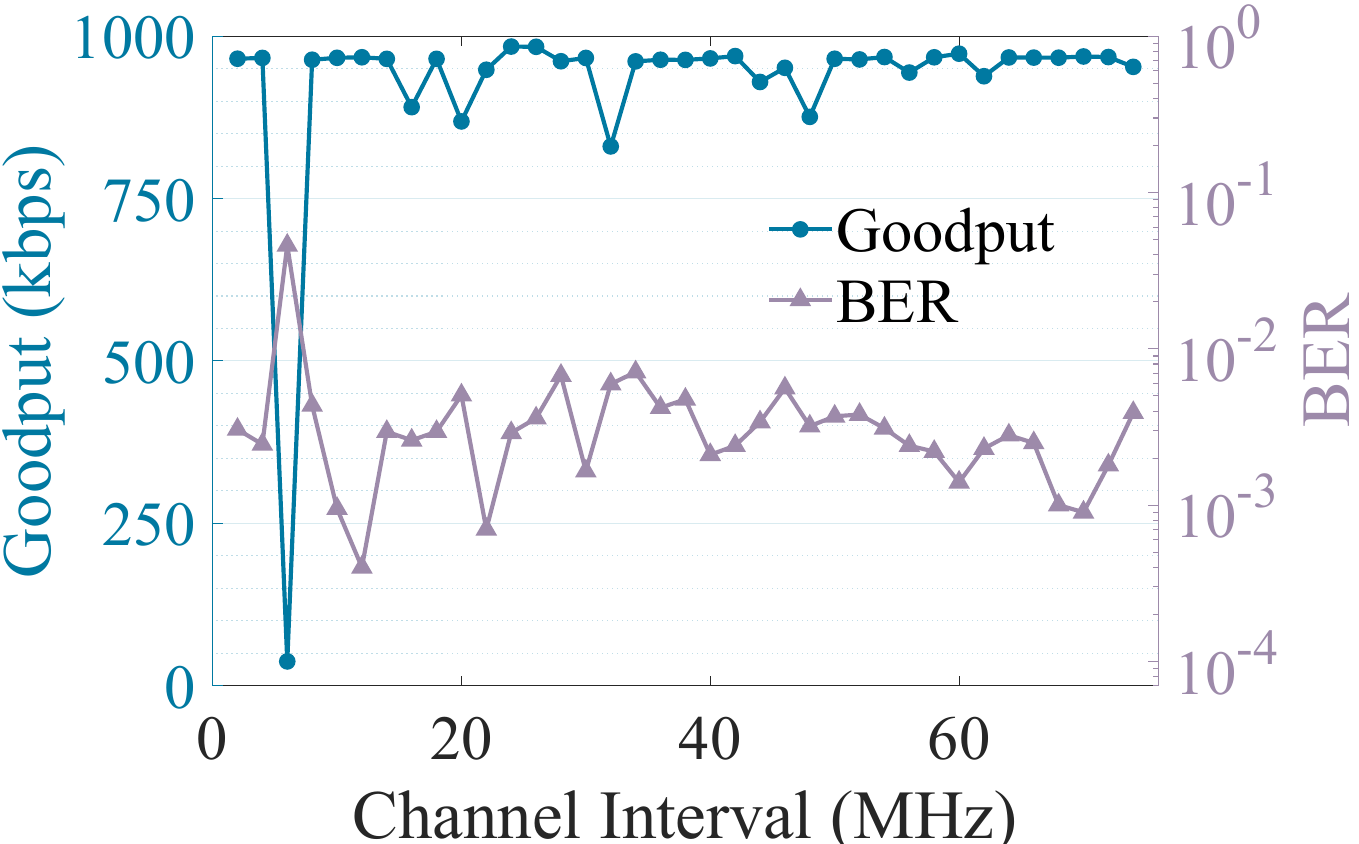}
          }
         \caption{ Impact of frequency shift between two \\ \sysname tags.}
         \label{fig:ConnectionInterval}
        \end{minipage}
        \begin{minipage}{0.47\linewidth} 
           \subfigure[LE 1M Mode.]{
			\includegraphics[width=0.47\linewidth]{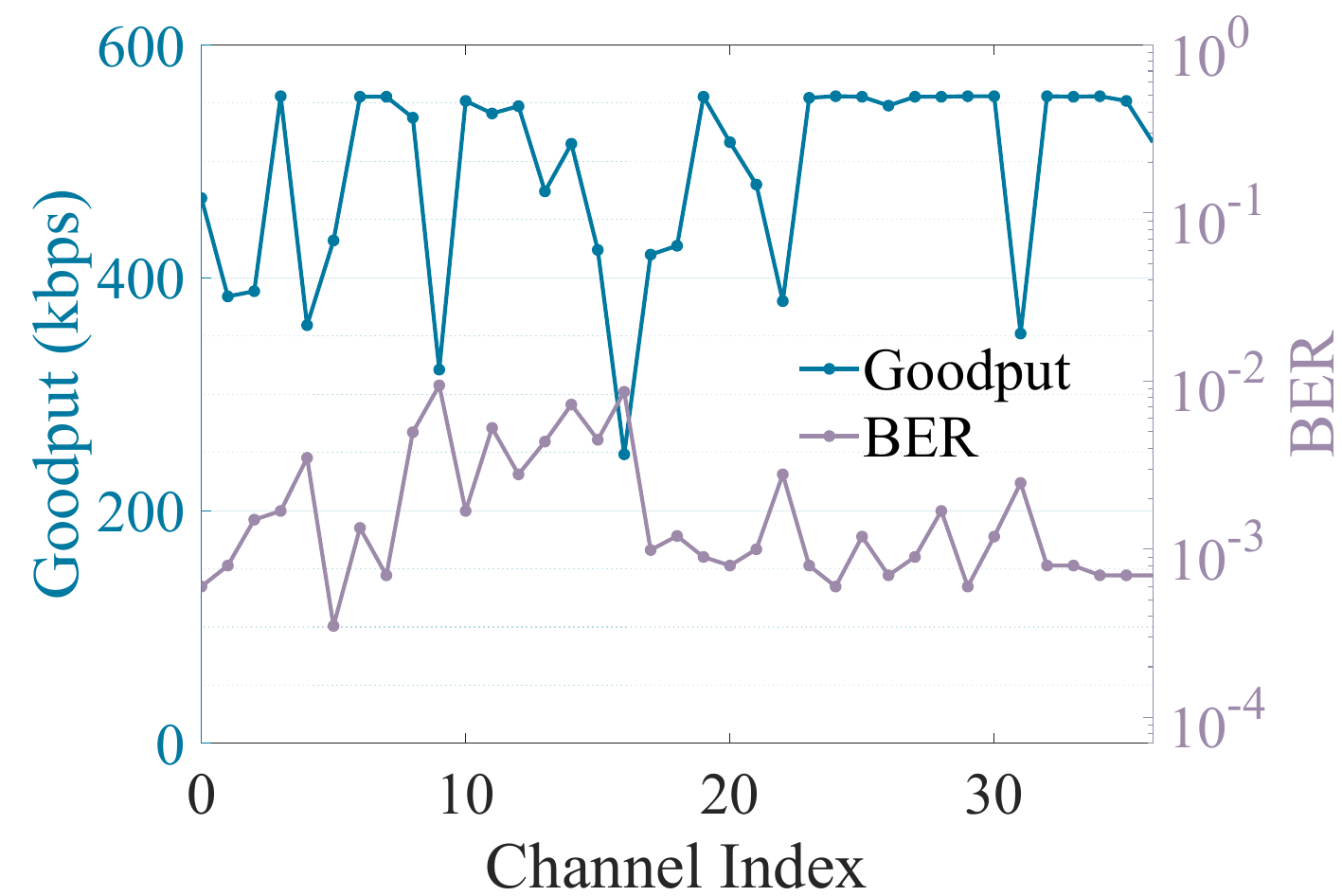}
           }
           \subfigure[LE 2M Mode.]{
			\includegraphics[width=0.47\linewidth]{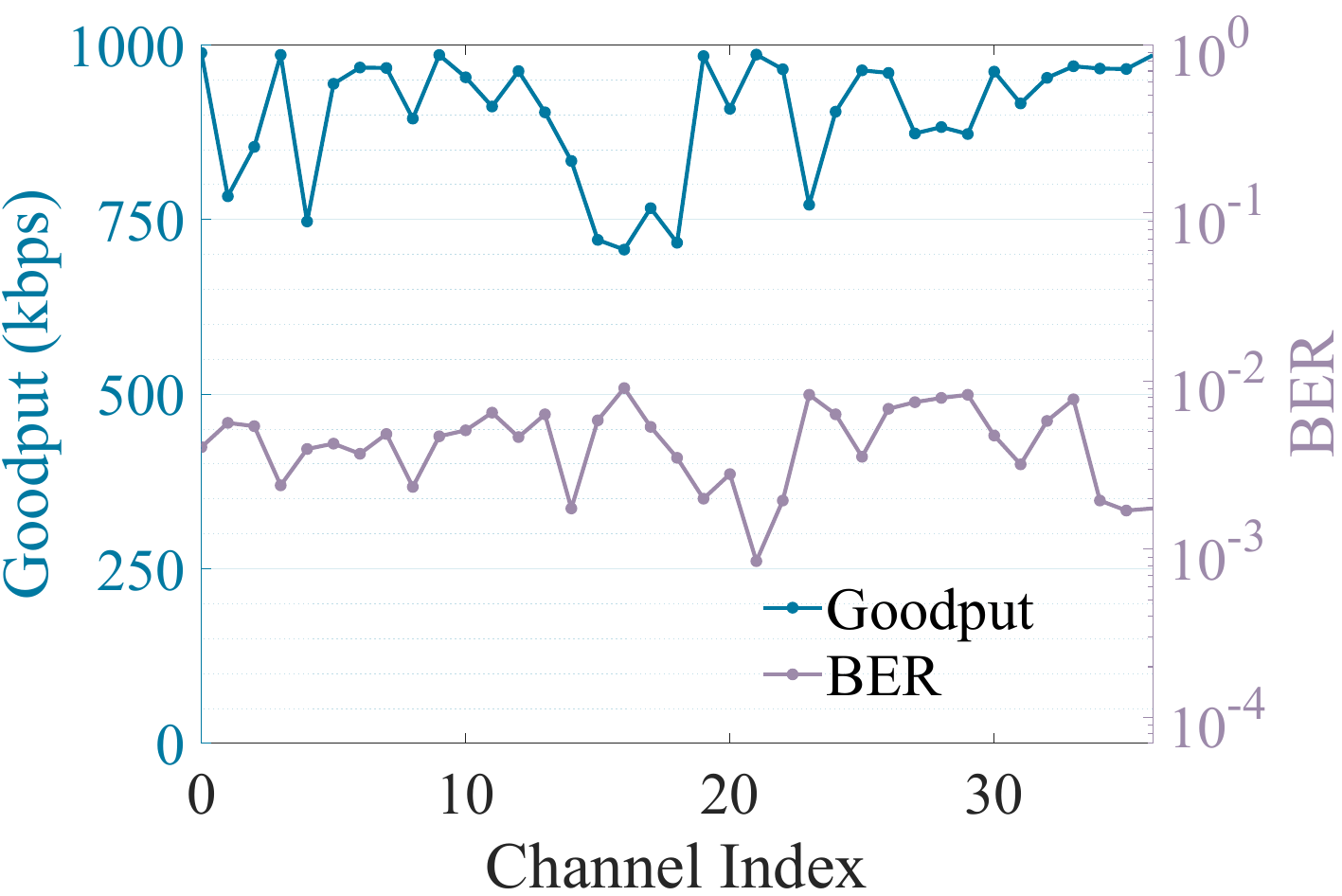}
           }
         \caption{Performance of a \sysname tag in different BLE channels.}
         \label{fig:ChannelIndex}
        \end{minipage}
         \vspace{-0.4cm}
\end{figure*}

\begin{figure}[t]
        \centering
        \setlength{\abovecaptionskip}{0.cm}
        \setlength{\belowcaptionskip}{0.cm}
           \subfigure[LE 1M Mode.]{
	    \includegraphics[width=0.47\linewidth]{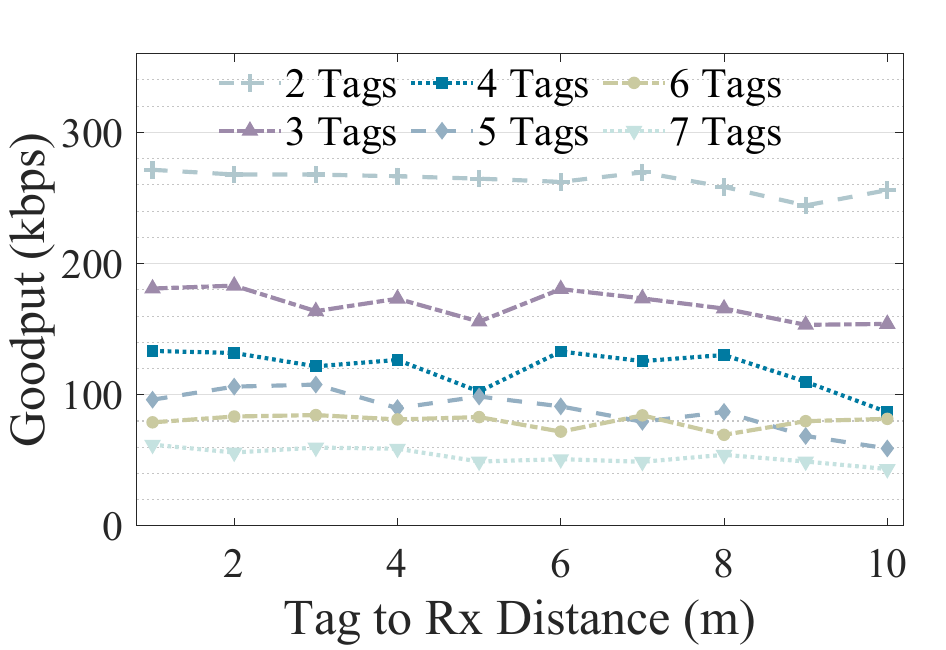}
          }
            \subfigure[LE 2M Mode.]{
			\includegraphics[width=0.47\linewidth]{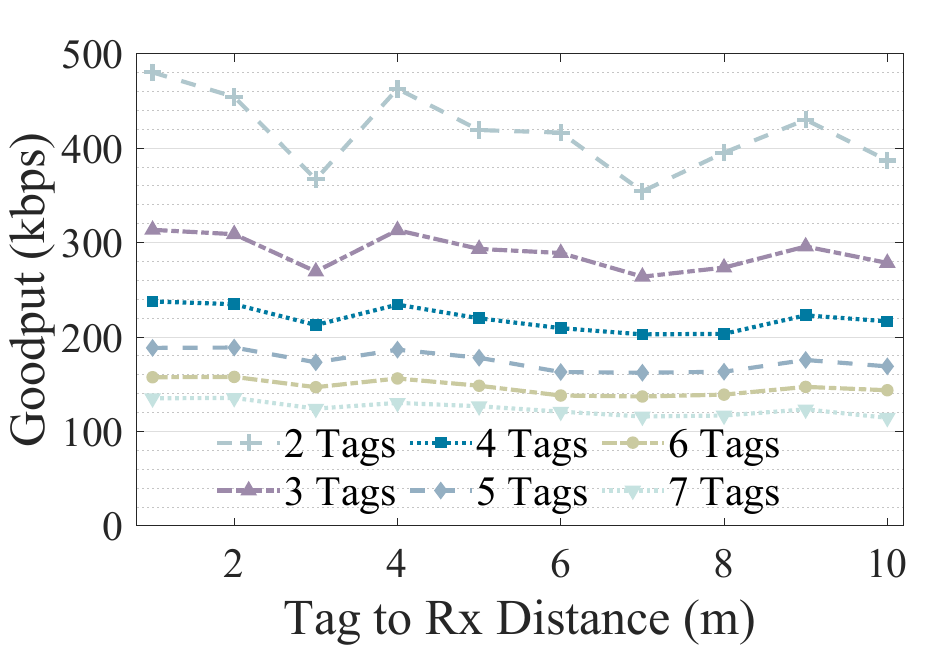}
          }
         \caption{ Average goodput of multiple tags with one excitation source.}
         \label{fig:ConnectionSingle}   
         \vspace{-0.3cm}
\end{figure} 

\vspace{-0.3cm}
\subsection{Connection Performance}
A significant advantage of \sysname is that it is the first BLE backscatter system capable of establishing commodity BLE connections with unmodified BLE transceivers, rather than one-shot communication. 
In this subsection, we conduct experiments to evaluate the performance in establishing and maintaining the BLE connections.

\textbf{Connection establishment performance.} To evaluate the successful connection establishment rates with respect to distance changes, we configure the \sysname tag to send advertising packets periodically, while the receiver listens and sends connection requests.
We consider the connection to be established upon a successful handshake in the advertising channel and receipt of a data packet from the \sysname tag by the receiver.
In contrast, the connection fails if it cannot be established within 6 consecutive connection events, as per the BLE protocol~\cite{BLESpecification}.
The connection-establishing process is repeated over 1000 times at each location. \fig~\ref{fig:Establishment_Maintance}~(a) shows the experimental results, which confirm that \sysname achieves a success rate of more than 99. 9\% in establishing BLE connections.

\textbf{Connection maintenance performance.} We also conduct experiments to evaluate the connection maintenance performance of \sysname tags. We manually set a 10-second connection maintenance target, and the connection event is configured as 50~ms. Similarly, a maintenance failure occurs if 6 consecutive connection events are lost. We repeat the 10-second connection maintenance experiments over 100 times at each distance. As plotted in Figure~\ref{fig:Establishment_Maintance}~(b), PassiveBLE tag can achieve a successful connection maintenance rate at 100\% within 14 meters. Then it degrades to below 33\% when the distance exceeds 17 meters. The success rate of connection maintenance is lower than the connection establishment rate under the same distance, which meets expectations.
Under the same packet loss rates, BLE connection establishment exhibits higher success probability because it only requires at least one successful handshake packet within six attempts. In contrast, connection maintenance involves prolonged data transmission, where the extended operational duration statistically increases exposure to rare six-packet loss streaks, despite equal per-packet failure probabilities. The connection maintenance faces a higher probabilistic risk due to longer exposure time to channel variability.

\textbf{Impact of the frequency hopping.} \label{Sec:channelDiversity}
Here we conduct a micro-benchmark experiment to evaluate the communication performance with/without the FHSS assistant on the excitation source. The \sysname tags is 0.5~m away from the excitation source and the receiver is 5~m away from the tag. We first configure the excitation source to send in fixed channels. In each channel, we collect 1000 BLE data packets generated by the tag to calculate the goodput and BER in that channel. The results in all of the BLE data channels that we tested one by one are plotted in \fig~\ref{fig:ChannelIndex}. The goodput and BER fluctuate roughly in different channels, where the goodput in the worst channel is less than half of the goodput in a good channel, e.g., the 240~kbps in channel 17 and 540~kbps in channel 24 as shown in \fig~\ref{fig:ChannelIndex}(a). 
We then test the performance of FHSS with randomly selected channels. In each channel, we collect 30000 packets in both of the modes to calculate the goodput and BER. Overall, the goodput is 902~kbps and 490~kbps in LE 2M and LE 1M mode, which is 93\% of the best case.

\textbf{Impact of frequency shift.}
We also conduct experiments to evaluate the impact of interference between two BLE tags that have different frequency shifts. We configure an excitation source to generate carriers at channel 19. Then configure two synchronized tags to generate BLE data packets with the carriers. The frequency interval of these two tags is configured ranging from 0~MHz, i.e. the same channel, to 80~MHz, i.e. over the whole BLE channels. In each frequency interval, we generate 1000 BLE data packets and use two BLE transceivers to collect these packets and calculate the average goodput and BER in these frequency intervals.   
As plotted in \fig\ref{fig:ConnectionInterval}, the frequency interference between two tags with different frequency shifts, results in fluctuating goodput and BER. But overall, \sysname achieves over 400~kbps goodput and $\leq$1\% BER, which suffices for typical BLE applications.

\textbf{Impact of multiple tags on one excitation source.}
In this subsection, we evaluate the performance of \sysname in maintaining multiple BLE connections with tags. As addressed in \S~\ref{Design}, the BLE connection scheduler manages multiple connections using time division duplex methods~\cite{BLESpecification}. 
To evaluate the performance of \sysname in such scenarios, we conduct experiments with multiple \sysname tags connected to one BLE transceiver with one excitation source, and the duration of each tag’s connection is configured to be $N\times50$~ms. We set 10 test locations with uplink distances ranging from 1~m to 10~m, and at each location, we record 5~minutes of data packets to evaluate the goodput. After all of the tags have successfully established the connection with the transceivers, the transceiver starts to collect the BLE packets and estimates the goodput. Results are plotted in \fig~\ref{fig:ConnectionSingle}, the average goodput of each tag drops rapidly than expected, which may caused by the processing delay. Overall, the average goodput for one tag can stay above 39~kbps, which is sufficient for logistic applications~\cite{zhang2017freerider}.

 \vspace{-0.3cm}
\section{Related Work}\label{related}

\textbf{BLE-based backscatter.}
As one of the most common ultra-low-power IoT communication methods, BLE backscatter~\cite{zhang2017freerider,InterScatter,zhang2020reliable,zhang2021commodity,EAScatter,jiang2023bidirectional,jiang2023dances,huang2024bitalign,reyes2024zeroscatter} has garnered substantial attention in recent years. In particular,
FreeRider~\cite{zhang2017freerider} and InterScatter~\cite{InterScatter} open up new opportunities of ambient IoT networking by scattering widely-adopted Bluetooth signals for ultra-low-power communications. Delving further, RBLE~\cite{zhang2020reliable}, IBLE~\cite{zhang2021commodity,jiang2023dances}, and EAScatter~\cite{EAScatter,huang2024bitalign} have dedicated substantial efforts to improve the reliability, compatibility, and goodput to enable practical BLE backscatter applications. 
Different from these systems, \sysname takes the first step to generate standard-compatible BLE data packets and establish fully commodity-compatible BLE data connections with unmodified BLE devices such as smartphones. 

\textbf{Low power demodulator.} 
Recently, the research community has shifted the focus of backscatter designs to new RF demodulators~\cite{guo2022saiyan,song2023mumote,dunna2021syncscatter,rostami2021mixiq} to replace the conventional envelope-detection-based demodulator. 
These demodulators are specifically designed for LoRa or WiFi signals, either relying on the frequency-increasing feature of chirp signals~\cite{guo2022saiyan,song2023mumote} or requiring multiple WiFi subcarriers~\cite{rostami2021mixiq}. SyncScatter~\cite{dunna2021syncscatter} designs and implements an outstanding wake-up radio chip for WiFi backscatter, which achieves 150~ns accuracy through a novel two-stage hierarchical envelope comparing methods. However, SyncScatter needs three WiFi packets (about $432\mu$s) to wake up tags, which is not suitable for our system due to the longer intervals of packets in the BLE protocol.
Departure from these approaches, \sysname demodulates the RF signals by comparing the signal with a delayed copy of itself, which meets the need for short delay and high accuracy in standard BLE communications.  

\textbf{Distributed Processing.} Distributed processing technologies~\cite{meng2024processor,li2020internet,katti2006xors, zhang2011dual} in the networking area, such as processor sharing~\cite{meng2024processor,li2020internet} of backscatter and network coding~\cite{katti2006xors, zhang2011dual}, pioneer the practice of offloading the burden from resource-limited devices to reduce power consumption or lower multicast traffic and improve network throughput. This insight inspires us to design a distributed coding scheme that offloads certain channel coding operations from tags to the excitation source. However, the XOR operation in \sysname differs technically from those systems, in which distributed processes occur in the baseband with logical circuits, while \sysname constructs XOR operations directly with RF signals.

\vspace{-0.3cm}
\section{Discussion}
\label{sec_discussion}
\textbf{Communication Distance.} The PassiveBLE system can achieve a maximum distance of about 17 meters with hundred-kbps-level goodput, which meets the requirements for short-range communication (e.g., wearable electronics, in-vehicle sensor networks, and logistics). The communication range still requires improvement compared to commodity active BLE chips, thereby limiting its usage in other BLE scenarios requiring longer ranges (e.g., device localization). The range is constrained primarily by the sensitivity (about -30 dBm) of the current synchronization circuit implementation. 
We believe that implement the system with tunnel diodes~\cite{dong2023gpsmirror,varshney2019tunnelscatter} or use $\mu W$ level LNA IC designs~\cite{taris201160muw} can further extend the communication distance.

\textbf{BLE security.} Using advertising packets for data uploading will expose the backscatter system to security risks, including eavesdropping, spoofing and data leakage, etc. Because the advertising packets are open to any BLE devices with publicly accessible rules. PassiveBLE system can enhance the security performance by leveraging BLE data packets for communication since the data packets are hard to eavesdrop with dynamic coding rules and seeds.

\vspace{-0.2cm}
\section{Conclusion}\label{conclusion}
We have designed and validated \sysnamenospace, which takes the first step to establish fully compatible and authentic BLE connections between passive backscatter tags and unmodified commodity BLE devices. We explore the design space of the low-power coherent-detection-based synchronization circuit, the XOR-based distributed coding scheme, as well as the excitation source-aided connection scheduler in backscatter communications. By extending backscatter communications to commodity BLE connections over data channels, \sysname can accommodate a wide range of BLE-suitable use cases.
\balance

\vspace{-0.3cm}
\section*{Acknowledgment}
We thank the anonymous reviewers and our shepherd, Yi-chao Chen, for their valuable comments. This research is supported in part by the National Natural Science Foundation of China with Grant 62471194, 62425207, Research Funds for the Central Universities, RGC under Contract CERG 16204523, AoE/E-601/22-R.

\balance
\bibliographystyle{ACM-Reference-Format}
\bibliography{ref}

\end{document}